\documentclass[twocolumn,preprintnumbers,prd,superscriptaddress,nofootinbib,floatfix,showpacs,showkeys]{revtex4-2}

\usepackage{amsmath}
\usepackage{amsfonts}
\usepackage{amssymb}
\usepackage{graphicx}
\usepackage{color}
\usepackage{xcolor}
\usepackage{hyperref}
\usepackage{booktabs}
\usepackage{hyperref}
\usepackage{cleveref}
\usepackage{braket}
\usepackage{mathtools}
\usepackage[rightcaption]{sidecap}
\usepackage{soul}
\usepackage{enumitem}
\usepackage{comment}
\usepackage{physics}
\usepackage{subfigure}
\usepackage{dsfont}
\usepackage{makecell}

\definecolor{vividviolet}{rgb}{0.62, 0.0, 1.0}
\definecolor{amaranth}{rgb}{0.9, 0.17, 0.31}
\definecolor{palatinateblue}{rgb}{0.15, 0.23, 0.89}
\definecolor{brightpink}{rgb}{1.0, 0.0, 0.5}
\definecolor{cornflowerblue}{rgb}{0.39, 0.58, 0.93}
\definecolor{deepcarminepink}{rgb}{0.94, 0.19, 0.22}
\definecolor{radicalred}{rgb}{1.0, 0.21, 0.37}

\hypersetup{ linktoc=all,
    colorlinks, linkcolor={palatinateblue},
    citecolor={brightpink}, urlcolor={amaranth}
}

\usepackage{tikz,xcolor}

\definecolor{lime}{HTML}{A6CE39}
\DeclareRobustCommand{\orcidicon}{%
	\begin{tikzpicture}
	\draw[lime, fill=lime] (0,0)
	circle [radius=0.16]
	node[white] {{\fontfamily{qag}\selectfont \tiny ID}};
	\draw[white, fill=white] (-0.0625,0.095)
	circle [radius=0.007];
	\end{tikzpicture}
	\hspace{-2mm}
}
\foreach \x in {A, ..., Z}{%
	\expandafter\xdef\csname orcid\x\endcsname{\noexpand\href{https://orcid.org/\csname orcidauthor\x\endcsname}{\noexpand\orcidicon}}
}

\begin{document}
%%%%%%%%%%%%%%%%%%%%%%%%%%%%%%%%%%%%%%%%%%%%%%%%%%
%%%%%%%%%%%%%%%%%%%%%%%%%%%%%%%%%%%%%%%%%%%%%%%%%%

\title{How cosmological expansion affects communication between distant quantum systems}

\author{Alessio Lapponi\orcidA{}}
\email{alessio.lapponi-ssm@unina.it}
\affiliation{Scuola Superiore Meridionale, Largo San Marcellino 10, 80138 Napoli, Italy.}
\affiliation{
Istituto Nazionale di Fisica Nucleare, Sezione di Napoli,
Complesso Universitario di Monte S. Angelo, Via Cintia Edificio 6, 80126 Napoli, Italy.}

\author{Orlando Luongo\orcidC{}}
\email{orlando.luongo@unicam.it}
\affiliation{Al-Farabi Kazakh National University, Al-Farabi av. 71, 050040 Almaty, Kazakhstan.}
\affiliation{School of Science and Technology, University of Camerino, Via Madonna delle Carceri 9, 62032 Camerino, Italy.}
\affiliation{SUNY Polytechnic Institute, 13502 Utica, New York, USA.}
\affiliation{Istituto Nazionale di Fisica Nucleare (INFN), Sezione di Perugia, Via A.~Pascoli, 06123 Perugia, Italy.}
\affiliation{INAF - Osservatorio Astronomico di Brera, Milano, Italy.}

\author{Stefano Mancini\orcidD{}}
\email{stefano.mancini@unicam.it}
\affiliation{School of Science and Technology, University of Camerino, Via Madonna delle Carceri 9, 62032 Camerino, Italy.}
\affiliation{Istituto Nazionale di Fisica Nucleare (INFN), Sezione di Perugia, Via A.~Pascoli, 06123 Perugia, Italy.}

\begin{abstract}
A quantum communication protocol between harmonic oscillator detectors, interacting with a quantum field, is developed in a cosmological expanding background. The aim is to see if the quantum effects arising in an expanding universe, such as the cosmological particle production, could facilitate the communication between two distant parts or if they provide an additive noisy effect. By considering a perfect cosmic fluid, the resulting expansion turns out to increase the classical capacity of the protocol. This increasing occurs for all the cosmological expansions unless the latter is sharpened just before the receiver's detector interacts with the field. Moreover, the classical capacity turns out to be sensible to the barotropic parameter $w$ of the perfect fluid and to the coupling between the field and the scalar curvature $\xi$. As a consequence, by performing this protocol, one can achieve information about the cosmological dynamics and its coupling with a background quantum field.
\end{abstract}

\vspace{0.5cm}

\maketitle
\tableofcontents

\section{Introduction}

Reconciling quantum mechanics with general relativity represents a deep challenge for the actual comprehension of  theoretical physics  \cite{DeWitt1967,Amelino_Camelia_2013,Ashtekar_2021}. Indeed, achieving a successful unification of these two theories would resolve fundamental inconsistencies, offering insights into the nature of fundamental interactions  \cite{1970RSPSA.314..529H,1979grec.conf..790W,Marletto2017WhyWN,Bongs2023}.

Despite considerable progresses, a fully consistent and experimentally confirmed theory of quantum gravity remains  elusive \cite{Carney_2019}. The pursuit of this theory continues to be a central focus in theoretical physics, with ongoing research exploring various approaches and potential experimental validations \cite{PRXQuantum.4.010320,Fuchs2024,Liang2024}.

A first attempt toward quantum gravity is provided by \textit{quantum field theory in curved spacetime} \cite{parker_toms_2009,Hollands_2015}, where gravity is treated semiclassical, namely it employs regimes where gravity is not too strong. Even though quantum gravity is thought to work in strong gravity regimes, quantum field theory in curved spacetimes predicts outstanding phenomena, such as the emission of radiation from horizons \cite{Hawking:1975vcx,Bhattacharya_2018} or from dynamical spacetime backgrounds \cite{NDBirrell_1980,Ford_2021,Good_2021}. Among them, the \textit{cosmological particle production} plays a prominent role in the understanding of the origin and evolution of our universe \cite{Ford1987,Ford_2021}. In particular, the particles produced by the universe expansion could be responsible for the origin of dark matter \cite{Herring_2020}, cosmic rays \cite{Dick_2006} and tensor perturbations of the cosmic microwave background \cite{Giovannini_2020}, as they could play a major role in baryogenesis \cite{Lima_2016}.

In particular, to comprehend the nature of particles produced by dynamical  spacetime backgrounds, it is essential to disclose which potential is associated with  detecting the produced particles. To this end, \textit{Unruh-DeWitt detectors} are typically considered \cite{Unruh1976,DeWitt}, representing  theoretical scenarios in which,  particle detectors imaged as quantum systems could interact with an external field. These systems have proven effective in achieving Unruh radiation  \cite{Unruh1976,Unruh1984,Schlicht_2004} and in understanding the radiation experienced by an observer following a generic worldline \cite{Perche_2022}. Further, directly observing particle production induced by gravitational fields in a laboratory has been explored using rotating particle detectors\footnote{Alongside analog gravity systems \cite{Jacquet_2020}, Unruh-DeWitt detectors are regarded as a promising approach for the direct observation of quantum effects induced by gravity.} \cite{D_Bunney_2023,Bunney_2024}.

Unruh-DeWitt models of particle detectors have recently been utilized to model Wi-Fi communication via quantum systems \cite{Cliche_2010,Landulfo_2016,Jonsson_2018,Tjoa_2022,Lapponi_2023,lapponi2024making}. Compared to classical systems, quantum devices may enhance the rate of classical message communication and enable the transmission of quantum messages \cite{mancini2019quantum}. The communication protocol involves a pair of particle detectors interacting through a quantum field. The properties of the resulting quantum channel have been determined non-perturbatively for both qubits \cite{Landulfo_2016,Tjoa_2022} and bosonic communication \cite{Lapponi_2023}. Recently, the feasibility of reliable Wi-Fi communication of quantum messages in a $(3+1)D$ spacetime has been demonstrated by leveraging the dynamics of the detectors \cite{lapponi2024making}.

The role of gravitational fields in this communication protocol remains unexplored. It is important to determine whether the effects predicted by quantum field theory in curved spacetime could enhance communication capabilities or pose an additional obstacle to reliable communication. The latter situation is proved to occur when considering information carried by the field's single modes\footnote{Specifically, it has been demonstrated that cosmological particle production degrades classical and quantum information stored in qubit states \cite{Mancini_2014} and bosonic states \cite{capozziello2024preservingquantuminformationfq}.} \cite{Gianfelici_2017,Good_2021}.

Motivated by these considerations, we here investigate how Wi-Fi communication of bosonic states is affected by cosmological expansion. To do so, we utilize a pair of one-dimensional harmonic oscillators interacting with the field only in a particular time. The classical capacity of the channel is shown to be sensitive to the cosmological expansion and to the coupling between the field and curvature. This demonstrates the potential to extract information about the dynamics of the universe through this communication protocol. Additionally, the measurement of the field-curvature coupling could be achieved, as it was shown that the channel capacity deviates from that in flat spacetime as the coupling with the scalar curvature, $\xi$, departs from its conformal value, i.e.,  $\xi=1/6$.

The paper is structured as follows: in Sec.~\ref{sec: Hamiltonian} we define the Hamiltonian of the physical system and compute its Heisenberg evolution by representing it with Guassian states. In Sec.~\ref{sec communication protocol} we define a quantum channel between the two detectors, whose properties and capacities can be studied non-perturbatively. In Sec.~\ref{sec: deltalike} we assume a rapid interaction between field and detector, finding the optimal parameters of the detectors to maximize the communication capabilities of the channel. Finally, in Sec.~\ref{sec: Cosmological expansion} we see how an accelerating cosmological expansion affects the classical capacity of the communication protocol, focusing on the expansion given by a perfect cosmological fluid\footnote{Through the paper, Planck units ($c=\hbar=8\pi G=1$) are considered.}.

\section{Particle detectors and their dynamics}\label{sec: Hamiltonian}

We consider two static particle detectors, named $A$ and $B$, respectively, interacting with a massless scalar quantum field $\hat{\Phi}$ and undergoing a cosmological expansion. To single out the background spacetime, adopting the cosmological principle, we assume the Friedmann-Robertson-Lemaitre-Walker (FRLW) line element, say
\begin{equation}\label{FLRW metric}
    ds^2=dt^2-a^2(t)(d\mathbf{x}\cdot d\mathbf{x})\,,
\end{equation}
written in Cartesian coordinates, where $a(t)$ represents the scale factor, depending upon the cosmic time, $t$. The positions of the detectors' center of mass, i.e. $\boldsymbol{\xi}_A$ and $\boldsymbol{\xi}_B$, by using the coordinates \eqref{FLRW metric}, is given by
\begin{equation}\label{detectors' position}
    \boldsymbol{\xi}_A=\mathbf{d}\,;\quad\boldsymbol{\xi}_B=\mathbf{0}\,.
\end{equation}
then, $d=|\mathbf{d}|$ is the conformal distance between the detectors.

\subsection{Hamiltonian}
The complete Hamiltonian of the system can be written as
\begin{equation}\label{complete Hamiltonian}
    \hat{H}_{tot}=\sum_{i=A,B}\left(\hat{H}_i+\hat{H}_I^i\right)+\hat{H}_\Phi\,,
\end{equation}
where   $\hat{H}_i$ is the Hamiltonian of the physical system representing the detector $i$, $\hat{H}_I^i$ is the Hamiltonian representing the interaction between the detector $i$ and the scalar field and, finally, $\hat{H}_\Phi$ is the Hamiltonian of the field.

As particle detectors, we consider a pair of non-relativistic one-dimensional harmonic oscillators, labelled by $i=A,B$, whose Hamiltonian reads
\begin{equation}\label{harmonic oscillator proper time}
    \hat{H}_i=\omega_i\left(\frac{1}{2}+\hat{a}_i^\dagger \hat{a}_i\right)\,,
\end{equation}
where $\omega_i$ is the frequency of the oscillator $i$ - or its \textit{energy gap}.

The interaction between the detector, $i$, and the scalar field is defined by coupling the field operator, $\hat{\Phi}$, to an observable of the detector $i$, usually called \textit{moment operator} of the detector, chosen to be
\begin{equation}\label{moment operator definition}
    \hat{q}_i=\frac{1}{\sqrt{2m_i\omega_i}}\left(\hat{a}_i^\dagger+\hat{a}_i\right)\,,
\end{equation}
where $m_i$ is the mass of the $ith$ oscillator.

The interacting Hamiltonian density can be written as
\begin{equation}\label{Interaction Hamiltonian Density}
\hat{h}_{I}^i(t,\mathbf{x})=f_i(\mathbf{x},t)\hat{q}_i(t)\otimes\hat{\Phi}(\mathbf{x},t)\,.
\end{equation}
In Eq.~\eqref{Interaction Hamiltonian Density}, the function $f_i(\mathbf{x},t)$ establishes how the field-detector interaction is distributed in space and time. We assume $f_i(\mathbf{x},t)=\lambda(t)\tilde{f}_i(\mathbf{x},t)$, where:
\begin{itemize}
    \item[-] $\tilde{f}_i(\mathbf{x},t)$, called \textit{smearing function}, indicates the position of the detector $i$ and its spatial distribution. The following normalization condition must be valid for each $t$
    \begin{equation}\label{normalization condition}
        \int_{\Sigma_{t}}\tilde{f}_i(\mathbf{x},t)\sqrt{\overline{g}(t,\mathbf{x})}d\mathbf{x}=1\,,
    \end{equation}
    where $\Sigma_{t}$ is the Cauchy surface $t=\text{const}$ and $\overline{g}=a^6(t)$ is the determinant of the spatial part the metric tensor.
    \item[-] $\lambda_i(t)$ is the \textit{switching-in function}, giving the strength of the interaction and how it turns on (and off) in time.
\end{itemize}
The interacting Hamiltonian is obtained by integrating Eq.~\eqref{Interaction Hamiltonian Density} in the time slice,  $t=\text{const}$, i.e.\small
\begin{align}
    H_I^i(t)&=\lambda(t)\int_{\Sigma_{t}}\tilde{f}_i(\mathbf{x},t)\hat{q}_i(t)\otimes \hat{\Phi}(t,\mathbf{x})\sqrt{-g(\mathbf{x},t)}d\mathbf{x}\nonumber\\&=\lambda_i(t)\hat{q_i}(t)\otimes \hat{\Phi}_{f_i}(t)\,,\label{interaction Hamiltonian}
\end{align}\normalsize
where $g(\mathbf{x},t)=\det g_{\mu\nu}(\mathbf{x},t)=-a(t)^6$ and where we defined for simplicity \begin{equation}\hat{\Phi}_f(t)\coloneqq\int_{\Sigma_t}\tilde{f}(\mathbf{x},t)\hat{\Phi}(\mathbf{x},t)\sqrt{-g(\mathbf{x},t)}d\mathbf{x}\,,
\end{equation}
called \textit{smeared field operator}.

Strictly speaking, the function $\tilde{f}_i(\mathbf{x},t)$, giving the shape of the detector, might have a compact support, to emphasize the fact that the detector has a finite size. However, one can also choose a distribution $\tilde{f}_i(\mathbf{x},t)$ with infinite support provided that $\tilde{f}$ is negligible outside a finite region of $\Sigma_t$. In this case, one can always define an effective finite size for the detector $i$, called $L_i$.

As better explained in Appendix \ref{appendix: Fermi Bound}, to consider a non-relativistic detector $i$ in a relativistic framework leads to upper bound $L_i$ by a length $L_F$, called \textit{Fermi lenght}, depending on the trajectory of the detector $i$ and on the spacetime it lies on. In our case, it results (see Appendix \ref{appendix: Fermi Bound}) \begin{equation}\label{Fermi bound CE}
    L_F=\frac{a}{\sqrt{\dot{a}^2-\ddot{a}a}}=\sqrt{\frac{6}{R}}\,,
\end{equation}
where we denoted with an upper dot the derivative with respect to $t$ and with $R$ the Ricci scalar curvature.

Moreover, the finite size $L_i$ of the detector $i$ provides an ultraviolet cutoff $E_c\propto L_i^{-1}$ on the energy of the field modes the detector $i$ interacts with \cite{Schlicht_2004}. If a detector has an energy cutoff $E_c^i$ for the particles it can interact with, a realistic situation is the one where the detector itself cannot have an energy greater than $E_c^i$.

Consequently, for the detector energy $E_i\coloneqq\langle \hat{H}_i\rangle$ we impose the upper bound
\begin{equation}\label{energetic condition for the detector}
    E_i\le E_c^i\,.
\end{equation}
We shall see later on how this bound provides a maximum rate of classical and quantum information that the two detectors can exchange.

\subsection{Quantum Langevin equations}

By using the Hamiltonian \eqref{complete Hamiltonian}, we study the Heisenberg evolution of the system. From Eq.~\eqref{interaction Hamiltonian}, the operator $q_i$ interacts linearly with the smeared field operator $\Phi_{f_i}$. By following Ref.~\cite{Ford1988}, the operator $q_i$ of a single oscillator interacting linearly with a field evolves according to a quantum Langevin equation. In particular, the field operator plays the role of an operator valued random force acting on the oscillators. Therefore, having a pair of harmonic oscillators interacting with the field, the Heisenberg evolution of the moment operators of the detectors $q_i$ is given by coupled quantum Langevin equations \cite{Ford1988,Lapponi_2023}, i.e.
\begin{widetext}
\begin{align}\label{vectorial form of langevin equation}
    \left(\begin{matrix}
        \frac{d^2}{dt^2}+\omega_A^2&0\\0&\frac{d^2}{dt^2}+\omega_B^2
    \end{matrix}\right)\left(\begin{matrix}
        q_A\\q_B
    \end{matrix}\right)-\int_{-\infty}^{t}\left(\begin{matrix}
        \frac{1}{m_A}&0\\0&\frac{1}{m_B}
    \end{matrix}\right)\left(\begin{matrix}
        \chi_{AA}(t,s)&\chi_{AB}(t,s)\\\chi_{BA}(t,s)&\chi_{BB}(t,s)
    \end{matrix}\right)\left(\begin{matrix}
        q_A(s)\\q_B(s)
    \end{matrix}\right)ds=\left(\begin{matrix}
        \frac{\lambda_A(t)}{m_A}\Phi_{f_A}(t)\\\frac{\lambda_B(t)}{m_B}\Phi_{f_B}(t)
    \end{matrix}\right)\,,
\end{align}\normalsize
\end{widetext}
where we defined the \textit{dissipation kernel}\small
\begin{equation}\label{dissipation kernel}
    \chi_{ij}(t,s)\coloneqq i\theta(t-s)\lambda_i(t)\lambda_j(t)\bra{\Phi}\left[\Phi_{f_i}(t),\Phi_{f_j}(s)\right]\ket{\Phi}\,,
\end{equation}\normalsize
with $\ket{\Phi}$ representing the initial state of the field, defined explicitly later on. Causality is respected. In fact, if the two detectors, localized by the smearings $\tilde{f}_A$ and $\tilde{f}_B$ are causally disconnected, then the off-diagonal elements of the dissipation kernel \eqref{dissipation kernel} are null. Consequently, the quantum Langevin equations \eqref{vectorial form of langevin equation} decouples and then the two detectors do not affect each other.

For later purposes, it is convenient to study the evolution of the dimensionless operator $\hat{Q}_i=\frac{a_i+a_i^\dagger}{\sqrt{2}}$, related to $\hat{q}_i$ from Eq.~\eqref{moment operator definition} through $\hat{Q}_i=\sqrt{m_i\omega_i}\hat{q}_i$. The coupled quantum Langevin equation \eqref{vectorial form of langevin equation} in terms of $\hat{Q}_i$, becomes
\begin{align}\label{compact QLE for Q}
    &\ddot{\mathbf{Q}}(t)+\Omega^2\mathbf{Q}(t)-\sqrt{\frac{\Omega}{\mathbb{M}}}\int_{-\infty}^{t}\chi(t,r)(\sqrt{\mathbb{M}\Omega})^{-1}\mathbf{Q}(r)dr\nonumber\\&=\sqrt{\frac{\Omega}{\mathbb{M}}}\boldsymbol{\varphi}(t)\,,
\end{align}
where $\mathbf{Q}\equiv(\hat{Q}_A,\hat{Q}_B)$, $\mathbb{M}\equiv\text{diag}(m_A,m_B)$, $\Omega\equiv\text{diag}(\omega_A,\omega_B)$, $\{\chi\}_{ij}=\chi_{ij}$ and \begin{equation}\boldsymbol{\varphi}(t)\coloneqq(\lambda_A(t)\Phi_{f_A}(t),\lambda_B(t)\Phi_{f_B}(t))\,.\end{equation}
As shown in Refs.~\cite{Lapponi_2023,lapponi2024making}, we can solve Eq.~\eqref{compact QLE for Q} by means of a Green function matrix
\begin{equation}
\mathbb{G}=\left(\begin{matrix}G_{AA}(t,s)&G_{AB}(t,s)\\G_{BA}(t,s)&G_{BB}(t,s)\end{matrix}\right),
\end{equation}
solution of the homogeneous form of the Langevin equation, Eq. \eqref{compact QLE for Q}, i.e.,\small
\begin{align}\label{homogeneous quantum langevin equation}
    &\ddot{\mathbb{G}}(t,s)+\Omega^2\mathbb{G}(t,s)-\sqrt{\frac{\Omega}{\mathbb{M}}}\int_{-\infty}^{t}\chi(t,r)(\sqrt{\mathbb{M}\Omega})^{-1}\mathbb{G}(r,s)dr\nonumber\\&=\delta(t-s)\mathbb{I}\,.
\end{align}\normalsize
By imposing a causal Green function matrix, i.e., $\mathbb{G}(t\le s)=0$, we have $\mathbb{G}(t=s,s)=0$ and $\dot{\mathbb{G}}(t=s,s)=\mathbb{I}$. Hence, the time evolution for $\mathbf{Q}$ in terms of $\mathbb{G}$ reads\footnotesize
\begin{align}\label{moment operator evolution}
    \mathbf{Q}(t)=&\dot{\mathbb{G}}(t,s)\mathbf{Q}(s)+\mathbb{G}(t,s)\dot{\mathbf{Q}}(s)+\int_{s}^{t} \mathbb{G}(t,r)\sqrt{\frac{\Omega}{\mathbb{M}}}\boldsymbol{\varphi}(r)dr\,.
\end{align}\normalsize

\subsection{Gaussian state formalism}\label{sec: oscillator hamiltonian}
To the purpose of studying how the two detectors communicate, it is worth choosing a convenient initial state for them. Then, we consider the initial state of the two oscillators to be a \textit{two-modes bosonic Gaussian state} \cite{serafini2017quantum} - where each oscillator represents one mode. By defining, for $i=A,B$, the \textit{quadrature operators}
\begin{equation}\label{quadrature operators}
    \hat{Q}_i=\frac{\hat{a}_i+\hat{a}_i^\dagger}{\sqrt{2}}\,\quad\hat{P}_i=\frac{\hat{a}_i-\hat{a}_i^\dagger}{i\sqrt{2}}\,,
\end{equation}
the two-modes bosonic state is Gaussian if every product of three or more quadrature operators \eqref{quadrature operators} has zero expectation value. Therefore, a two-modes Gaussian state is represented only by:
\begin{itemize}
    \item[-] the \textit{first momentum vector}
\begin{equation}\label{first momentum vector}
        \mathbf{d}=\left(\langle \hat{Q}_A\rangle,\langle\hat{P}_A\rangle,\langle\hat{Q}_B\rangle,\langle\hat{P}_B\rangle\right)\,,
    \end{equation}
\item[-] the \textit{covariance matrix}
\begin{equation}\label{covariance matrix physical basis}
     \sigma=\left(\begin{array}{c|c}
       \sigma_{AA} & \sigma_{AB}\\\hline
        \sigma_{BA} & \sigma_{BB}
    \end{array}\right)\,,
\end{equation}
\end{itemize}
where, for $i,j=A,B$\footnotesize
\begin{equation}
    \sigma_{ij}=\frac{1}{2}\left(\begin{matrix}
        \langle\left\{Q_i,Q_j\right\}\rangle-\langle Q_i\rangle\langle Q_j\rangle&\langle\left\{Q_i,P_j\right\}\rangle-\langle Q_i\rangle\langle P_j\rangle\\\langle\left\{P_i,Q_j\right\}\rangle-\langle P_i\rangle\langle Q_j\rangle&\langle\left\{P_i,P_j\right\}\rangle-\langle P_i\rangle\langle P_j\rangle
    \end{matrix}\right)\,.\label{covariance matrix submatrices}
\end{equation}\normalsize

In Eq.~\eqref{covariance matrix physical basis}, the submatrix $\sigma_{ii}$, with $i=A,B$, represents a one-mode Gaussian state describing the state of the oscillator $i$. The off-diagonal submatrices $\sigma_{AB}=\sigma_{BA}^T$, instead, represent the correlations between the two detectors $A$ and $B$ - imposed to vanish before the detectors interact with the field.

All the entropic quantities of a Gaussian state are independent from the first momentum vector, Eq. \eqref{first momentum vector}. Hence, we can consider $\mathbf{d}=\mathbf{0}$ without loss of generality and simplify Eq.~\eqref{covariance matrix submatrices} to
\begin{equation}\label{Simplified submatrix}
    \sigma_{ij}=\frac{1}{2}\left(\begin{matrix}
        \langle\left\{Q_i,Q_j\right\}\rangle&\langle\left\{Q_i,P_j\right\}\rangle\\\langle\left\{P_i,Q_j\right\}\rangle&\langle\left\{P_i,P_j\right\}\rangle
    \end{matrix}\right)\,.
\end{equation}
Up to a unitary transformation, the covariance matrix, in Eq.~\eqref{Simplified submatrix}, of the single harmonic oscillator $i$ can be further simplified to
\begin{equation}\label{generic detector state}
    \sigma_{ii}=\left(\begin{matrix}
        \left(\frac{1}{2}+N_i\right)e^{l_i}&0\\0&\left(\frac{1}{2}+N_i\right)e^{-l_i}
    \end{matrix}\right)\,,
\end{equation}
where $N_i$ is the average number of entropic particles of the state and $l_i$ is the \textit{squeezing parameter}.

The \textit{Von Neumann entropy} of the state represented by $\sigma_{ii}$, in Eq.~\eqref{generic detector state}, reads
\begin{equation}\label{Von Neumann entropy input}
    \mathcal{S}(\sigma_{ii})=(N_{i}+1)\log(N_{i}+1)-N_{i}\log(N_{i})\,,
\end{equation}
where, by convention, with $\log$ we denote a base $2$ logarithm.

At this point, Eq.~\eqref{harmonic oscillator proper time} can be rewritten is terms of the quadrature operators, Eqs. \eqref{quadrature operators}, as
\begin{equation}
    H_i=\frac{\omega_i}{2}\left(\hat{Q}_i^2+\hat{P}_i^2\right)\,,
\end{equation}
In this way, the energy of the detector $i$ can be quantified as the expectation value
\begin{equation}\label{detector energy}
    E_i=\langle \hat{H}_i\rangle=\frac{\omega_i}{2}\text{Tr}\sigma_{ii}=\omega_i\left(\frac{1}{2}+N_i\right)\cosh{l_i}\,.
\end{equation}

Having represented the two oscillators' system via the covariance matrix \eqref{covariance matrix physical basis}, we can now study how this evolves via the interaction with the field.

From the definition of the quadrature operators in Eqs. \eqref{quadrature operators}, one can easily check that $\dot{\hat{Q}}_i=\omega_i\hat{P}_i$, so that Eq.~\eqref{moment operator evolution} can be rewritten as\footnotesize
\begin{align}\label{moment operator evolution for momentum}
    \mathbf{Q}(t)=&\dot{\mathbb{G}}(t,s)\mathbf{Q}(s)+\mathbb{G}(t,s)\Omega\mathbf{P}(s)+\int_{s}^{t} \mathbb{G}(t,r)\sqrt{\frac{\Omega}{\mathbb{M}}}\boldsymbol{\varphi}(r)dr\,,
\end{align}\normalsize
where $\mathbf{P}=(\hat{P}_A,\hat{P}_B)$.

At this stage, we can also compute the evolution of the quadrature operator $\hat{P}_i$ by applying a time derivative to Eq.~\eqref{moment operator evolution for momentum} and multiplying by $\mathbb{F}^{-1}(t_B)\Omega^{-1}$ from the left
\begin{align}\label{momentum evolution}
    \mathbf{P}(t)=&\Omega^{-1}\ddot{\mathbb{G}}(t,s)\mathbf{Q}(s)+\Omega^{-1}\dot{\mathbb{G}}(t,s)\Omega\mathbf{P}(s)\nonumber\\&+\Omega^{-1}\int_{s}^{t}\dot{\mathbb{G}}(t,r)\sqrt{\frac{\Omega}{\mathbb{M}}}\boldsymbol{\varphi}(r)dr\,.
\end{align}\normalsize
The evolution of the operators $\hat{Q}_i$ and $\hat{P}_i$, from Eqs.
\eqref{moment operator evolution for momentum} and \eqref{momentum evolution}, allows us to compute the corresponding covariance matrix $\sigma$ dynamics, from the time $s$ up to the time $t$, obtaining \cite{Lapponi_2023,lapponi2024making}
\begin{equation}\label{two-modes Gaussian channel}
    \sigma(t)=\mathbb{T}_2\sigma(s)\mathbb{T}_2^T+\mathbb{N}_2\,,
\end{equation}
where
\begin{equation}\label{complete transmission matrix}
    \mathbb{T}_2=P\left(\begin{array}{c|c}
       \dot{\mathbb{G}}(t,s) & \mathbb{G}(t,s)\Omega\\\hline
        \Omega^{-1}\ddot{\mathbb{G}}(t,s) & \Omega^{-1}\dot{\mathbb{G}}(t,s)\Omega
    \end{array}\right)P\,,
\end{equation}
\begin{equation}\label{complete noise matric}
    \mathbf{N}=P\left(\begin{array}{c|c}
    \mathbf{N}_{QQ}&\mathbf{N}_{QP}\\\hline\mathbf{N}_{QP}^T&\mathbf{N}_{PP}\end{array}\right)P\,,
\end{equation}\normalsize
with
\begin{equation}
    P\coloneqq\left(\begin{matrix}
        1&0&0&0\\0&0&1&0\\0&1&0&0\\0&0&0&1
    \end{matrix}\right),
\end{equation}

\begin{equation}
    \mathbf{N}_{QQ}=\int_s^t\int_s^t \mathbb{G}(t,r)\sqrt{\frac{\Omega}{\mathbb{M}}}\nu(r,r')
\sqrt{\frac{\Omega}{\mathbb{M}}}\mathbb{G}^{T}(t,r')drdr'\,;
\end{equation}
\begin{equation}
    \mathbf{N}_{QP}=\int_s^t\int_s^t \mathbb{G}(t,r)\sqrt{\frac{\Omega}{\mathbb{M}}}\nu(r,r')
\sqrt{\frac{\Omega}{\mathbb{M}}}\dot{\mathbb{G}}^{T}(t,r')\Omega^{-1}drdr'\,;
\end{equation}
\begin{equation}
    \mathbf{N}_{PP}=\int_s^t\int_s^t\Omega^{-1}\dot{\mathbb{G}}(t,r)\sqrt{\frac{\Omega}{\mathbb{M}}}\nu(r,r')
\sqrt{\frac{\Omega}{\mathbb{M}}}\dot{\mathbb{G}}^{T}(t,r')\Omega^{-1}drdr'\,.
\end{equation}\normalsize
Here, we defined the \textit{noise kernel} $\nu(t,t')$ as
\begin{align}
    &\nu_{ij}(t,t')\coloneqq\{\nu(t,t')\}_{ij}=\frac{\lambda_i(t)\lambda_j(t')}{2}\langle\left\{\hat{\Phi}_{f_i}(t),\hat{\Phi}_{f_j}(t')\right\}\rangle\,.\label{noise kernel}
\end{align}

\section{Communication channel between two detectors}\label{sec communication protocol}

Now, we can easily define a quantum communication channel between the detectors. The protocol consists in the sender preparing a state (where information is encoded) of the detector $A$ at the time $s$ and let the interaction with the field occur. This should hopefully transfer the state at later times to the detector $B$ - once it also interacts with the field. We wonder how much information about the oscillator $A$ at the time $s$ can be achieved from the receiver's detector $B$ state at the time $t$. The former is represented by the submatrix $\sigma_{AA}$ of the covariance matrix $\sigma(s)$ from Eq.~\eqref{covariance matrix physical basis}. The latter by the submatrix $\sigma_{BB}$ of $\sigma(t)$. We can figure out a communication channel map
\begin{equation}
    \mathcal{N}:\sigma_{in}=\sigma_{AA}(s)\mapsto\sigma_{BB}(t)=\sigma_{out}\,.
\end{equation}
Since we know how the system evolves from Eq.~\eqref{two-modes Gaussian channel}, using Eqs. \eqref{complete transmission matrix} and \eqref{complete noise matric}, we get
\begin{equation}\label{input-output relation}
    \sigma_{out}=\mathbb{T}\sigma_{in}\mathbb{T}^T+\mathbb{N}\,.
\end{equation}
The matrices $\mathbb{T}$ and $\mathbb{N}$ are, respectively:
\begin{equation}\label{transmissivity matrix}
    \mathbb{T}=\left(\begin{matrix}
        \dot{G}_{BA}(t,s)&G_{BA}(t,s)\omega_A\\\frac{\ddot{G}_{BA}(t,s)}{\omega_B}&\dot{G}_{BA}(t,s)\frac{\omega_A}{\omega_B}
    \end{matrix}\right)\,;
\end{equation}\small
\begin{equation}\label{noise matrix}
    \mathbb{N}=\left(\begin{matrix}
        \dot{G}_{BB}&G_{BB}\omega_B\\\ddot{G}_{BB}\omega_B^{-1}&\dot{G}_{BB}
    \end{matrix}\right)\sigma_{BB}(s)\left(\begin{matrix}
        \dot{G}_{BB}&\ddot{G}_{BB}\omega_B^{-1}\\G_{BB}\omega_B&\dot{G}_{BB}
    \end{matrix}\right)+\mathbb{N}'_B\,,
\end{equation}\normalsize
where $\mathbb{N}'_B=\left(\begin{matrix}
        N_{11}&N_{12}\\N_{12}&N_{22}
    \end{matrix}\right)$ with\begin{widetext}\small
\begin{align}
    N_{11}=&\frac{\omega_A}{m_A}\int_s^t\int_s^tG_{BA}(t,r)\nu_{AA}(r,r')G_{BA}(t,r')drdr'+\sqrt{\frac{\omega_A\omega_B}{m_Am_B}}\int_s^t\int_s^t G_{BB}(t,r)\nu_{BA}(r,r')G_{BA}(t,r')drdr'\nonumber\\&+\sqrt{\frac{\omega_A\omega_B}{m_Am_B}}\int_s^t\int_s^tG_{BA}(t,r)\nu_{AB}(r,r')G_{BB}(t,r')drdr'+\frac{\omega_B}{m_B}\int_s^t\int_s^tG_{BB}(t,r)\nu_{BB}(r,r')G_{BB}(t,r')drdr'\label{N11}\,;
\end{align}
\begin{align}
    N_{12}=&\frac{\omega_A}{m_A\omega_B}\int_s^t\int_s^tG_{BA}(t,r)\nu_{AA}(r,r')\dot{G}_{BA}(t,r')drdr'+\sqrt{\frac{\omega_A}{m_Am_B\omega_B}}\int_s^t\int_s^tG_{BB}(t,r)\nu_{BA}(r,r')\dot{G}_{BA}(t,r')drdr'\nonumber\\&+\sqrt{\frac{\omega_A}{m_Am_B\omega_B}}\int_s^t\int_s^tG_{BA}(t,r)\nu_{AB}(r,r')\dot{G}_{BB}(t,r')drdr'+m_B^{-1}\int_s^t\int_s^tG_{BB}(t,r)\nu_{BB}(r,r')\dot{G}_{BB}(t,r')drdr'\label{N12}\,;
\end{align}
\begin{align}
    N_{22}=&\frac{\omega_A}{m_A\omega_B}\int_s^t\int_s^t\dot{G}_{BA}(t,r)\nu_{AA}(r,r')\dot{G}_{BA}(t,r')drdr'+\sqrt{\frac{\omega_A}{m_Am_B\omega_B^3}}\int_s^t\int_s^t\dot{G}_{BB}(t,r)\nu_{BA}(r,r')\dot{G}_{BA}(t,r')drdr'\nonumber\\&+\sqrt{\frac{\omega_A}{m_Am_B\omega_B^3}}\int_s^t\int_s^t\dot{G}_{BA}(t,r)\nu_{AB}(r,r')G_{BB}(t,r')drdr'+\frac{1}{m_B\omega_B}\int_s^t\int_s^t\dot{G}_{BB}(t,r)\nu_{BB}(r,r')\dot{G}_{BB}(t,r')drdr'\label{N22}\,.
\end{align}
\normalsize\end{widetext}
The first term of the matrix $\mathbb{N}$, from Eq.~\eqref{noise matrix}, represents the evolution of detector $B$'s initial state $\sigma_{BB}(s)$, while the second term, $\mathbb{N}'_B$, is a contribute given by the interaction of detector $B$ with the field. For later purposes, it is also useful to know how the state of the oscillator $A$ behaves after the interaction with the field, namely, at the time $t$, supposing the detector $B$ has no longer influence to the detector $A$, we have\footnotesize
\begin{equation}
    \sigma_{AA}(t)\sim\left(\begin{matrix}
        \dot{G}_{AA}&G_{AA}\omega_A\\\ddot{G}_{AA}\omega_A^{-1}&\dot{G}_{AA}
    \end{matrix}\right)\sigma_{AA}(s)\left(\begin{matrix}
        \dot{G}_{AA}&\ddot{G}_{AA}\omega_A^{-1}\\G_{AA}\omega_A&\dot{G}_{AA}
    \end{matrix}\right)+\mathbb{N}'_A\,,\label{Alice's detector state}
\end{equation}\normalsize
where $\mathbb{N}'_A$ is equal to $\mathbb{N}'_B$ by exchanging the labels $A$ and $B$.

If the output of the channel $\mathcal{N}$, i.e. $\sigma_{out}$ can be written in terms of the input $\sigma_{in}$ as in Eq.~\eqref{input-output relation}, with $\det\mathbb{N}\ge\frac{1}{2}|1-\det \mathbb{T}|$, then the channel $\mathcal{N}$ is a \textit{one-mode Gaussian channel} \cite{Holevo2007}. Each one-mode Gaussian channel $\mathcal{N}$ can be reduced to its canonical form $\mathcal{N}_c$ by applying two unitary operations $U_{in}$ and $U_{out}$ to the input and the output of $\mathcal{N}$, respectively. Accordingly, we have
\begin{equation}\label{Canonical form mapping}
    \mathcal{N}_c=U_{out}\circ\mathcal{N}\circ U_{in}:\sigma_{in}\mapsto\tau\sigma_{in}+\sqrt{W}\mathbb{I}\,.
\end{equation}
The parameter $\tau\equiv\det\mathbb{T}$ in Eq.~\eqref{Canonical form mapping}, named \textit{transmissivity} of the channel $\mathcal{N}$, indicates the fraction of the input state's amplitude effectively present to the output. The parameter $W\equiv\det\mathbb{N}$, instead, indicates the additive noise created by the channel.

In particular, the \textit{average number of noisy particles} detected is
\begin{equation}\label{additive noise}
    \overline{n}\coloneqq\begin{cases}
        &\frac{\sqrt{W}}{|1-\tau|}-\frac{1}{2}\,\quad\text{if}\quad \tau\ne1\,;\\
        &\sqrt{W}\,\quad\text{otherwise}\,.
    \end{cases}
\end{equation}
A one-mode Gaussian channel is then completely characterized by $\tau$ and $W$.

Moreover, we say that the quantum channel $\mathcal{N}$ is \textit{entanglement-breaking} if every $n$-mode entangled state, input of $\mathcal{N}^{\otimes n}$ is mapped into a separable $n$-mode state. A one-mode Gaussian channel, characterized by $\tau$ and $W$, is entanglement-breaking if and only if \cite{Holevo2008ent}
\begin{equation}\label{Entanglement-breaking condition}
    W\ge\frac{1}{2}(1+\tau)\,.
\end{equation}

Finally, by considering $\tau$ and $W$, it is possible to compute the \textit{capacities} of the one-mode Gaussian channel $\mathcal{N}$, quantifying the capabilities of the channel to communicate information \cite{holevo1999evaluating,Pilyavets_2012,Br_dler_2015}. In particular, the \textit{classical capacity} (\textit{quantum capacity}) of a channel $\mathcal{N}$ is the maximum rate of classical information (quantum information) that the channel $\mathcal{N}$ can reliably transmit.

In the following sections, we focus exclusively on the classical capacity, denoted with $C$. The motivation lies on the fact that the quantum capacity is expected to be zero as a consequence of the no-cloning theorem, since both the interaction with the field and the background spacetime are isotropic, see e.g. Refs.~\cite{Jonsson_2018,lapponi2024making} for a more complete explanation.

Indeed, we later prove that the channel $\mathcal{N}$ we consider is entanglement-breaking, so that its quantum capacity is zero and its classical capacity $C$ can be analytically computed up the energy bound $E_c^A$ of the input state \cite{Shor_2002}. In particular, in Appendix \ref{appendix: capacities}, we compute the classical capacity $C$ as
\begin{equation}\label{Classical capacity of the channel}
    C(\tau,W)=h\left(\frac{E_c^A}{\omega_A}\tau+\sqrt{W}\right)-h\left(\frac{\tau}{2}+\sqrt{W}\right)\,,
\end{equation}
where $\omega_A$ its frequency of the input mode.

\section{Rapid interaction between field and detectors}\label{sec: deltalike}

We now study the communication protocol with the interaction between the detectors and the field occurring only at a particular time, i.e., a \textit{rapid interaction}. This kind of interaction has recently been proven to provide exact solutions for the communication properties of Wi-Fi communication channel between detectors \cite{Tjoa_2022,lapponi2024making}.

The time at which the detector $i$ interacts with the field is denoted with $t_I^i$. In so doing, the switching-in functions of the detector $i$ can be written as $\lambda_i(t)=\lambda_i\delta(t-t_i^I)$. The spacetime smearing of the detectors $f_{i=A,B}$ are now written as
\begin{equation}\label{deltalike spacetime smearing}
    f_i(\mathbf{x}_i,t_i)=\lambda_i\delta(t_i-t_I^i)\tilde{f}(\mathbf{x}_i)\,.
\end{equation}
In this case, the elements $\chi_{AA}(t,s)$ and $\chi_{BB}(t,s)$ of the dissipation kernel \eqref{dissipation kernel} are non-null only when $t=s=t_I^A$ and $t=s=t_I^B$, respectively. However, those elements become expectation value operators, commuting with themselves, implying $\chi_{AA}=\chi_{BB}=0$.

The homogeneous quantum Langevin equation, in Eq. \eqref{homogeneous quantum langevin equation}, can be therefore simplified as
\begin{widetext}
\begin{equation}\label{QLE delta like true}
    \begin{cases}
        &\ddot{G}_{AA}(t,s)+\omega_A^2 G_{AA}(t,s)-\sqrt{\frac{\omega_A}{\omega_Bm_Am_B}}\int_{-\infty}^t\chi_{AB}(t,r)G_{BA}(r,s)dr=0\,,\\
        &\ddot{G}_{AB}(t,s)+\omega_A^2 G_{AB}(t,s)-\sqrt{\frac{\omega_A}{\omega_Bm_Am_B}}\int_{-\infty}^t\chi_{AB}(t,r)G_{BB}(r,s)dr=0\,,\\
        &\ddot{G}_{BA}(t,s)+\omega_B^2 G_{BA}(t,s)-\sqrt{\frac{\omega_B}{\omega_Am_Am_B}}\int_{-\infty}^t\chi_{BA}(t,r)G_{AA}(r,s)dr=0\,,\\
        &\ddot{G}_{BB}(t,s)+\omega_B^2 G_{BB}(t,s)-\sqrt{\frac{\omega_B}{\omega_Am_Am_B}}\int_{-\infty}^t\chi_{BA}(t,r)G_{AB}(r,s)dr=0\,.
    \end{cases}
\end{equation}\normalsize\end{widetext}
with boundary conditions $G_{ij}(t\to s^+,s)=0$ and $\dot{G}_{ij}(t\to s^+,s)=\delta_{ij}$.

Since by hypothesis the detector $A$ communicates its state to the detector $B$, we require $t_I^B>t_I^A$, for guaranteeing causality.

Hence, from Eq.~\eqref{dissipation kernel}, we can immediately see that\footnote{This means that the interaction of the oscillator $B$ with the field does not affect the state of the detector $A$, validating the hypothesis to obtain Eq.~\eqref{Alice's detector state}.} $\chi_{AB}^A(t,s)\sim0$. For the dissipation kernel element $\chi_{BA}$, instead, we have\footnotesize
\begin{align}
    \chi_{BA}(t,s)&=i\lambda_A\lambda_B\delta(t-t_I^B)\delta(s-t_I^A)\bra{\Phi}\left[\hat{\Phi}_{f_A}(t_I^A),\hat{\Phi}_{f_B}(t_I^B)\right]\ket{\Phi}\nonumber\\
    &=\lambda_A\lambda_B\delta(t-t_I^B)\delta(s-(t_I^A)_B)I(t_I^A,t_I^B)\,,\label{dissipation kernel delta like}
\end{align}\normalsize
where for the sake of simplicity we called
\begin{equation}\label{I integral def}
    I(t_I^A,t_I^B)\coloneqq i\bra{\Phi}\left[\hat{\Phi}_{f_A}(t_I^A),\hat{\Phi}_{f_B}(t_I^B)\right]\ket{\Phi}\,.
\end{equation}
The first and second relations of the homogeneous quantum Langevin equation, Eqs. \eqref{QLE delta like true}, become
\begin{equation}
    \ddot{G}_{Ai}+\omega_A^2G_{Ai}=0\,,
\end{equation}
whose solutions for $i=A,B$ are respectively
\begin{align}
    &G_{AA}(t,s)=\frac{\sin(\omega_A(t-s))}{\omega_A}\,,\label{GAA general}\\
    &G_{AB}(t,s)=0\,.
\end{align}
Since $G_{AB}=0$, then also the fourth of Eq.~\eqref{QLE delta like true} can be solved as
\begin{equation}\label{GBB general}
    G_{BB}(t,s)=\frac{\sin(\omega_B(t-s))}{\omega_B}\,.
\end{equation}
Finally, from the third relation in Eqs.~\eqref{QLE delta like true}, we obtain
\begin{align}\label{GBA general}
    G_{BA}(t,s)=&\theta(t-t_I^B)\sqrt{\frac{\omega_B}{\omega_Am_Am_B}}\frac{\lambda_A\lambda_B}{\omega_A\omega_B}I(t_I^B,t_I^A)\nonumber\\&\times\sin(\omega_B(t-t_I^B))\sin(\omega_A(t_I^A-s))\,.
\end{align}
The transmissivity $\tau$ can now be easily computed from the determinant of $\mathbb{T}$, defined in Eq.~\eqref{transmissivity matrix}, resulting into
\begin{align}
    \tau(t,s)&=\frac{\omega_A}{\omega_B}\left(\dot{G}_{BA}^2(t,s)-G_{BA}(t,s)\ddot{G}_{BA}(t,s)\right)\nonumber\\&=\frac{\Lambda_A\Lambda_B}{\omega_A^2}I^2(t_I^B,t_I^A)\sin^2(\omega_A(t_I^A-s))\,,\label{tau delta-like}
\end{align}
where we set $\Lambda_i\equiv\frac{\lambda_i^2}{m_i}$.

To compute the noise parameter $W$, from the determinant of the matrix $\mathbb{N}$ in Eq.~\eqref{noise matrix}, we compute the noise kernel elements \eqref{noise kernel} as
\begin{align}
    \nu_{ij}(t,s)&=\frac{\lambda_i\lambda_j}{2}\delta(t-t_I^i)\delta(s-t_I^j)\nonumber\\&\times\int d\mathbf{x}_i\int d\mathbf{x}_j\tilde{f}_i(\mathbf{x}_i)\tilde{f}_j(\mathbf{x}_j)\sqrt{g(\mathbf{x}_i,t_I^i)g(\mathbf{x}_j,t_I^j)}\nonumber\\&\times\bra{\Phi}\left\{\hat{\Phi}(\mathbf{x}_i,t_I^i),\hat{\Phi}(\mathbf{x}_j,t_I^j)\right\}\ket{\Phi}\,.\label{noise kernel element delta-like}
\end{align}
By using the Green function matrix elements \eqref{GAA general}, \eqref{GBB general} and \eqref{GBA general}, Eqs.~\eqref{N11}, \eqref{N12} and \eqref{N22} drastically simplify to
\begin{subequations}\label{N simplified}
    \begin{align}
    N_{11}&=\frac{\Lambda_B}{\omega_B}\sin^2(\omega_B(t-t_I^B))J_B(t_I^B)\,,\\
    N_{11}&=N_{12}=\frac{\Lambda_B}{\omega_B}\sin(\omega_B(t-t_I^B))\cos(\omega_B(t-t_I^B))J_B(t_I^B)\,,\\
    N_{22}&=\frac{\Lambda_B}{\omega_B}\cos^2(\omega_B(t-t_I^B))J_B(t_I^B)\,,
    \end{align}
\end{subequations}
where, for the sake of simplicity, we defined
\begin{equation}\label{J definition}
    J_i(t_I^i)\equiv \frac{1}{2}\bra{\Phi}\left\{\hat{\Phi}_{f_i}(t_I^i),\hat{\Phi}_{f_i}(t_I^i)\right\}\ket{\Phi}\,.
\end{equation}
At this stage, using Eqs.~\eqref{N simplified}, we can compute the parameter $W\equiv\det\mathbb{N}$ through Eq.~\eqref{noise matrix}. By considering a generic initial state $\sigma_{BB}(s)$ of the detector $B$ from Eq.~\eqref{generic detector state}, we get\small
\begin{align}
    &W\coloneqq\det\mathbb{N}=\left(\frac{1}{2}+N_B\right)^2+\frac{\Lambda_B}{\omega_B}J(t_I^B)\left(\frac{1}{2}+N_B\right)\nonumber\\&\times\left(e^{l_B}\cos^2(\omega_B(t_I^B-s))+e^{-l_B}\sin^2(\omega_B(t_I^B-s))\right)\,.\label{noise delta-like}
\end{align}\normalsize

\subsection{Optimizing the classical capacity}\label{ssec: maximization}

Looking at Eq.~\eqref{Classical capacity of the channel}, $C$ clearly increases with $\tau$ and decreases with $W$. Thus, we aim to find the optimal set of parameters $\omega_i,\Lambda_i,N_B,l_B$ maximizing the transmissivity, $\tau$, from Eq.~\eqref{tau delta-like} and minimizing $W$, from Eq.~\eqref{noise delta-like}.

To increase $C$, one might increase the parameters $\Lambda_A$ and $\Lambda_B$. However, those parameters cannot be increased arbitrarily due to the energy condition reported in Eq. \eqref{energetic condition for the detector}. Indeed, once the detector $i$ interacts with the field, the contribute $\mathbb{N}'_i$ appears in the covariance matrix $\sigma_{ii}$ - as indicated in Eqs.~\eqref{input-output relation} and \eqref{noise matrix} when $i=B$ and in Eq.~\eqref{Alice's detector state} when $i=A$. Hence, we can compute the energy absorbed by the detector $i$, when interacting with the field $E_I^i$ by means of Eq.~\eqref{detector energy}, yielding
\begin{equation}
    E_I^i=\frac{\omega_A}{2}\text{Tr}(\mathbb{N}'_i)=\frac{\Lambda_iJ_i(t_I^i)}{2}\,.
\end{equation}
However, the energy $E_I^i$ cannot be larger than the energy bound $E_c^i$ of the detector $i$. This means that the maximum value $\Lambda_i$ can have is
\begin{equation}\label{Interaction bound}
    \Lambda_i=2\frac{E_c^i}{J_i(t_I^I)}\,.
\end{equation}

Another physical limit is given by the Heisenberg principle, imposing an uncertainty $\Delta t_i^I$ on the interaction time $t_I^i$. As explained in details in Ref.~\cite{lapponi2024making}, a rapid interaction between distant detectors could be considered only if $\Delta t_i^I\ll |t_I^i-t_I^j|$, i.e. if the uncertainty of the interaction time is much smaller than the time needed to perform the communication protocol. Since $\Delta t_I^i\propto \frac{1}{E_i}$, being $E_i$ the energy of the detector $i$, and since $E_i\ge\frac{\omega_i}{2}$, then the protocol violates the Heisenberg principle if $\omega_i$ is chosen arbitrarily low. Hence, for $\omega_i$, we must have
\begin{equation}\label{frequency lower bound}
    \omega_i|t_I^i-t_I^j|\gg1\,.
\end{equation}
From now on, since $\frac{\omega_i}{2}\le E_c^i$, we conveniently write $\omega_i=2\alpha_iE_c^i$, where $\alpha_i$ is upper bounded by $1$. The condition \eqref{frequency lower bound} then becomes
\begin{equation}\label{Condition for alpha}
    \frac{1}{2E_c^i|t_I^i-t_I^j|}\ll\alpha_i\le1\,.
\end{equation}
It worth noticing that the left hand side of Eq.~\eqref{Condition for alpha} is necessarily much smaller than $1$, since $E_c^i\propto L_i$ and since we are considering a communication at distance.

At this point, to maximize the capacity, we increase $\Lambda_A$ and $\Lambda_B$ up to the bound given by Eq.~\eqref{Interaction bound}. In so doing, the transmissivity, $\tau$, and the noise, $W$, become respectively

\begin{subequations}
\begin{align}
    &\tau=\frac{1}{\alpha_A^2}\frac{E_c^B}{E_c^A}\frac{I^2(t_I^B,t_I^A)}{J_A(t_I^A)J_B(t_I^B)}\sin^2(\omega_A(t_I^A-s))\,,\label{maximum transmissivity}\\
    &W=\left(\frac{1}{2}+N_B\right)^2+\frac{1}{\alpha_B}\left(\frac{1}{2}+N_B\right)\times\nonumber\\&\times\left(e^{l_B}\cos^2(\omega_B(t_I^B-s))+e^{-l_B}\sin^2(\omega_B(t_I^B-s))\right)\,.\label{simplified noise}
\end{align}\normalsize
\end{subequations}

Accordingly, we can now minimize the noise, $W$, even by finding the optimal parameters,  $\alpha_B$, $n_B$ and $l_B$, while respecting the condition in Eq. \eqref{energetic condition for the detector}. This minimization occurs when\footnote{Other combinations of the parameters $\alpha_B$ and $l_B$ lead to the same minimized result for $W$ in Eq.~\eqref{minimized W} (namely $l_B=\ln(|\tan(\omega_B(t_I^B-s))|)$ and $\alpha_B=(\cosh(l_B))^{-1}$). However, if we consider this general result, whenever $\omega_B(t_I^B-s)=k\frac{\pi}{2}$ where $k\in\mathbb{Z}$, we have $\alpha\sim0$, forbidden from the condition $\alpha_i\gg(2d_i E_c^i)^{-1}$. Without loss of generality, we can consider the receiver preparing his state in the unsqueezed ground state $l_B=n_B=0$, and choose an energy gap $\omega_B=2E_c^B$ to have a minimization of $W$ regardless the phase $\omega_B(t_I^B-s)$.} $n_B=0$, $\alpha_B=1$ and $l_B=0$, leading to
\begin{equation}\label{minimized W}
    W=\frac{3}{4}\,.
\end{equation}

Last but not least, we notice that the transmissivity, $\tau$, in Eq.~\eqref{maximum transmissivity}, depends on $t_I^A-s$, i.e., on the time the sender awaits from the preparation of the state $s$ to the interaction with the field $t_I^A$. This dependence is also evident as the two detectors travel with different trajectories \cite{lapponi2024making}.

Thus, we consider the sender to wait a time $t_I^A-s=\frac{\pi}{2\omega_A}$, before interacting with the field - because of Eq.~\eqref{frequency lower bound}, this time should be no longer than communication time $t_I^B-t_I^A$. In this way, the transmissivity becomes
\begin{equation}\label{maximum transmissivity 2}
    \tau=\frac{1}{\alpha_A^2}\frac{E_c^B}{E_c^A}\frac{I^2(t_I^B,t_I^A)}{J_A(t_I^A)J_B(t_I^B)}\,.
\end{equation}
The maximized classical capacity, obtained by using Eqs.~\eqref{minimized W} into Eq.~\eqref{Classical capacity of the channel}, becomes
\begin{equation}\label{optimized classical capacity}
    C=h\left(\frac{\tau}{2\alpha_A}+\sqrt{\frac{3}{4}}\right)-h\left(\frac{\tau}{2}+\sqrt{\frac{3}{4}}\right)\,.
\end{equation}
Therefore, Eq.~\eqref{optimized classical capacity} gives the maximum rate of information that two distant harmonic oscillator detectors can reliably transmit. In the next section, we evaluate if this classical capacity is enhanced or decreased when considering a cosmological expansion.

\section{Communication during a cosmological expansion}\label{sec: Cosmological expansion}

In the coordinate system, $(t,\mathbf{x})$, used to characterize the FRW metric, in Eq.~\eqref{FLRW metric}, the detectors' center of masses are placed in $\boldsymbol{\xi}_A=\mathbf{d}$ and $\boldsymbol{\xi}_B=\mathbf{0}$ (see also Eq.~\eqref{detectors' position}).

For the smearing of the detectors $\tilde{f}_A(\mathbf{x},t)$ and $\tilde{f}_B(\mathbf{x},t)$, we consider a Lorentzian shape centered around the detectors' center of mass $\boldsymbol{\xi}_i$ which reads
\begin{equation}\label{Lorentzian shape}
    \tilde{f}_i(\mathbf{x},t)=\frac{1}{\pi a^3(t)}\frac{\epsilon}{\left(|\mathbf{x}-\boldsymbol{\xi}_i|^2+\epsilon^2\right)^2}\,,
\end{equation}
where $\epsilon$ can be seen as the detectors' \textit{effective size} \cite{Schlicht_2004,lapponi2024making}, and a normalization has been pursued through Eq.~\eqref{normalization condition}. \\

To compute the Fermi bound \eqref{Fermi bound CE}, one must consider the detectors' effective size in the detectors' proper frame. As discussed in Appendix \ref{appendix: Fermi Bound}, the local non-relativistic coordinates in the detector's proper frame are the Fermi-normal coordinates \eqref{Fermi normal coordinates}. By using those, the smearing \eqref{Lorentzian shape} becomes
\begin{equation}\label{Lorentzian shape Fermi coordinates}
    \tilde{f}_i(\mathbf{x}_i,t)=\frac{1}{\pi}\frac{a(t)\epsilon}{\left(\mathbf{x}_i\cdot\mathbf{x}_i+a^2(t)\epsilon^2\right)^2}\,.
\end{equation}
We then recognize $a(t)\epsilon$, scaling with the expanding universe, as the effective size of the detectors in their proper coordinates. The Fermi bound \eqref{Fermi bound CE} then implies
\begin{equation}\label{Fermi bound cosmological expansion}
    \epsilon^{-1}\gg a\sqrt{\frac{R}{6}}\,.
\end{equation}
Having the two detectors the same effective size, then for the energy cutoffs we have $E_c^A=E_c^B$, satisfying
\begin{equation}
    \epsilon E_c^i\ll1\,.
\end{equation}
As a consequence, the transmissivity \eqref{maximum transmissivity 2} becomes
\begin{equation}\label{maximum transmissivity 3}
    \tau=\frac{1}{\alpha_A^2}\frac{I^2(t_I^B,t_I^A)}{J_A(t_I^A)J_B(t_I^B)}\,,
\end{equation}
where $\epsilon/d\ll\alpha_A\le1$. The explicit computation of the transmissivity from Eq.~\eqref{maximum transmissivity 3} is reported in Appendix \ref{appendix: perturbation}. There, to obtain an explicit analytic solution for $\tau$, two approximation are performed:
\begin{enumerate}
    \item To consider a Minkowski vacuum as initial state of the field - as \textit{initial}, we mean at the time $t_I^A$ - we considered expansions where $\dot{a}(t_I^A)$ is negligible, so that the Riemann curvature is also negligible and the initial spacetime could be approximated as Minkowskian;
    \item To obtain solutions of the Klein-Gordon equation in the metric \eqref{FLRW metric}, we used a perturbation method developed in Ref.~\cite{Zeldovich:1971mw} and reported in details in the Appendix \ref{appendix: perturbation} that considers the expansion as a perturbation of the Minkowski case.
\end{enumerate}
By using the conformal time $\eta$ s.t. $a(\eta)d\eta=dt$ and by defining the conformal Hubble parameter $H_c(\eta)\coloneqq\frac{a'}{a^2}$ - where the prime $'$ indicates a derivative with respect to $\eta$ - the transmissivity \eqref{maximum transmissivity 3}, from the Appendix \ref{appendix: perturbation}, results to be
\begin{equation}\label{ maximum transmissivity cosmological expansion main text}
    \tau=\frac{16}{\alpha_A^2}\frac{\epsilon^2}{d^2}(1+(1-6\xi)F)\,,
\end{equation}
where
\begin{align}
    F\coloneqq\int_{\eta_I^A}^{\eta_I^B}\left(\frac{\pi\epsilon}{2}-\frac{4\epsilon^2(\eta_I^B-\eta)}{(\eta_I^B-\eta)^2+\epsilon^2}\right)a^2(\eta)H_c^2(\eta)d\eta\,.\label{factor F main text}
\end{align}
To determine whether the perturbation method is valid or not, in the Appendix \ref{appendix: perturbation} we also computed the maximum relative error we have on $\tau$ by considering only the first order perturbation theory. This relative error is
\begin{equation}
    \mathcal{E}_P\coloneqq 4\int_{\eta_I^A}^\eta\int_{\eta_I^A}^{\eta_1}(\eta-\eta_1)(\eta_1-\eta_2)U(\eta_1)U(\eta_2)d\eta_1d\eta_2\,,\label{Perturbation error main text}
\end{equation}
where
\begin{equation}\label{potential}
    U(\eta)\coloneqq \left(6\xi-1\right)\left(\frac{a'(\eta)}{a(\eta)}\right)^2=(6\xi-1)a^2(\eta)H_c^2(\eta)\,.
\end{equation}
Furthermore, a relative error $\mathcal{E}_F$ could be associated to the constraint \eqref{Fermi bound cosmological expansion} given by the Fermi bound, namely
\begin{equation}\label{Fermi bound error}
    \mathcal{E}_F=a\epsilon\sqrt{\frac{R}{6}}\,.
\end{equation}
In this way Eq.~\eqref{Fermi bound cosmological expansion} is equivalent to say $\mathcal{E}_F\ll1$.

The maximum relative error we have by evaluating $F$ from Eq.~\eqref{factor F main text} is then
\begin{equation}\label{Maximum relative error}
    \mathcal{E}=\sqrt{\mathcal{E}_P^2+\mathcal{E}_F^2}\,.
\end{equation}

In a Minkowski spacetime $F=0$, hence Eq.~\eqref{ maximum transmissivity cosmological expansion main text} reduces to
\begin{equation}\label{maximum transmissivity Minkowski}
    \tau_M=\frac{16}{\alpha_A^2}\frac{\epsilon^2}{d^2}\,.
\end{equation}
Since $\alpha_A\gg\epsilon/d$ from Eq. \eqref{Condition for alpha}, then $\tau_M\ll1$ from Eq.~\eqref{maximum transmissivity Minkowski}.

We are interested to know how an accelerated expansion affects the classical capacity of the protocol with respect to the Minskowski case, where no expansion occurs. To know that, we use $\Delta \tau$ to evaluate the relative increment of the transmissivity due to an accelerated expansion, i.e.
\begin{equation}
    \Delta\tau\coloneqq\frac{\tau-\tau_M}{\tau_M}=(1-6\xi)F\,.
\end{equation}

Since $\tau\sim \tau_M$ up to a perturbative term, we have $\tau\ll1$ as well. Then, from Eq.~\eqref{minimized W} we immediately see that the condition \eqref{Entanglement-breaking condition} is satisfied, confirming the assumption that the channel is entanglement-breaking. Henceforth, since $\tau\ll1$, the classical capacity \eqref{optimized classical capacity} can be approximated by \cite{Shor_2002}
\begin{align}
    C&\sim\log\left(\frac{\sqrt{3}+1}{\sqrt{3}-1}\right)\frac{1}{2}\left(\frac{1}{\alpha_A}-1\right)\tau\nonumber\\&\simeq 0.95\cdot\left(\frac{1}{\alpha_A}-1\right)\tau\label{Capacity}\,.
\end{align}
Henceforth, by considering a cosmological expansion, also for the classical capacity $C$ we have a relative increment
\begin{equation}\label{relative increasing of the capacity}
    \Delta C\equiv\frac{C-C_M}{C_M}=(1-6\xi)F\,.
\end{equation}

Finally, we can conclude that a cosmological expansion:
\begin{enumerate}
    \item increases the capabilities of two static particle detectors to communicate classical information if $\xi<1/6$ and $F>0$ or if $\xi>1/6$ and $F<0$;
    \item decreases them if $\xi<1/6$ and $F<0$ or if $\xi>1/6$ and $F>0$;
    \item leaves them completely unaltered in case of \textit{conformal coupling} $\xi=1/6$ or if $F=0$.
\end{enumerate}
The function $F$ is an integral of a positive definite function $H^2_c(\eta)$ weighted with a function
\begin{equation}\label{g Function}
    g(\eta):=\frac{\pi\epsilon}{2}-4\frac{\epsilon^2(\eta_I^B-\eta)}{(\eta_I^B-\eta)^2+\epsilon^2}\,.
\end{equation}
\begin{figure*}
    \centering
    \includegraphics[scale=1]{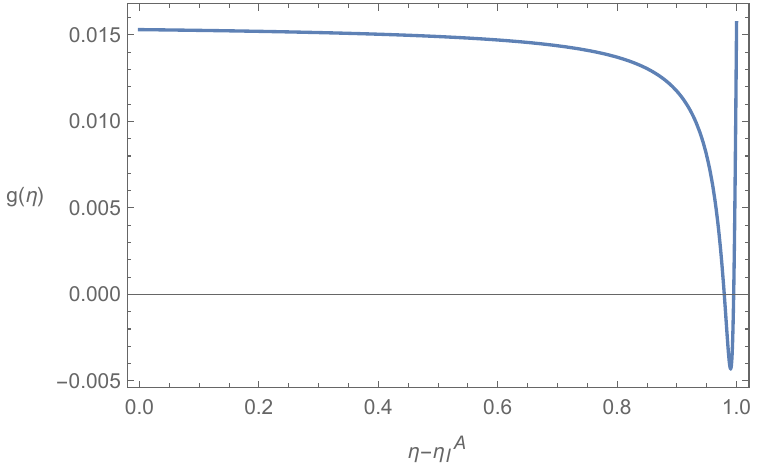}
    \caption{Weight function $g(\eta)$, in terms of $\eta$ (Eq.~\eqref{g Function}) needed for the integration of $H_c^2(\eta)$ to obtain the factor $F$ in Eq.~\eqref{factor F}. The parameters $\epsilon=0.01$ and $d=1$ were chosen.}
    \label{fig:gFunction}
\end{figure*}
Thus, the sign of $F$ can be estimated by studying the sign of $g(\eta)$, which is plotted in Fig.~\ref{fig:gFunction} for this purpose. We can see that the latter is always positive except in a small range. Actually, $g(x)$ is negative when
\begin{align}\label{range F negative}
    \frac{\eta_I^B-\eta}{\epsilon}
    &\in\left(\frac{4}{\pi}-\sqrt{\frac{16}{\pi^2}-1},\frac{4}{\pi}+\sqrt{\frac{16}{\pi^2}-1}\right)\nonumber\\
    &\simeq(0.49,2.06)\,.
\end{align}
However, this interval is very small - by a factor $\sim1.576\frac{\epsilon}{d}$ -
with respect to the interval $\left[\eta_I^A,\eta_I^B\right]$ where $\eta$ is integrated in Eq.~\eqref{factor F main text}. Then, the factor $F$ could be negative only in the specific situation in which $H_c^2(\eta)$ is really high in the range \eqref{range F negative} and negligible elsewhere. This happens if one has a sudden cosmological expansion at the range \eqref{range F negative} and stops afterwards.

Apart from this very specific and singular situation, we can say that $F$ is positive and then, that the classical capacity is increased by a cosmological expansion as long as $\xi<1/6$ (including the \textit{minimal coupling} $\xi=0$) and decreased by it if the coupling $\xi$ is greater than $1/6$.

\subsection{The case of Einstein-de Sitter Universe}\label{ssec: perfect fluid}

We now consider the specific example of a cosmological expansion given by a perfect fluid whose equation of state is $p=w\rho$ with a constant barotropic parameter $w$.\\

Solving the Einstein equations with Eq.~\eqref{FLRW metric}, in conformal time, we obtain
\begin{subequations}
    \begin{align}\label{Fried1}
&H_c^2= \frac{1}{3}\rho\,,\\
&\frac{H_c'}{a}+H_c^2  = -\frac{1}{6}\left(\rho+3p\right)\,.\label{Fried2}
    \end{align}
\end{subequations}
Coming the two relations above, the continuity equation holds,

\begin{equation}
    \frac{\rho'}{a}+3H_c(w+1)\rho=0,
\end{equation}
that can be recast to give
\begin{equation}\label{Friedmann Eq. perfect fluid}
    \frac{H_c'}{a}+\frac{3(1+w)}{2}H_c^2=0\,,
\end{equation}

Now, it is convenient to assume that one fluid dominates over the other species. Hence, the pressure and density, $p$ and $\rho$, respectively, are associated with a given equation of state, namely $w\equiv \frac{p}{\rho}$. This universe is called Einstein-de Sitter (EdS), for which Eq.~\eqref{Friedmann Eq. perfect fluid} can be solved exactly. By excluding the case $w=-1$, in the Appendix \ref{appendix: scale factor} we compute
\begin{equation}\label{conformal Hubble parameter perfect fluid}
    a(\eta)H_c(\eta)=\frac{1}{\frac{3w+1}{2}\eta+\eta_I^A}\,.
\end{equation}
It is worth remarking that $\dot{a}(t_I^A)$ was considered enough small to approximate the initial state of the field as the Minkowski vacuum. However, to be consistent with this choice, the flat case should be considered at every time $t$ involving $\dot{a}(t)<\dot{a}(t_I^A)$. Then, to see the non-negligible effects from cosmological expansion, we consider exclusively accelerated expansions, i.e., from Eq.~\eqref{scale factor perfect fluid cosmological time}, $w<-1/3$.

At this stage, the factor $F$ in Eq.~\eqref{factor F main text} can be analytically computed. Indeed, using
\begin{enumerate}
\item  the fact that $d=\eta_I^B-\eta_I^A\gg\epsilon$,
    \item the validity of first order perturbation theory,
    \item the Fermi bound,
\end{enumerate}
we infer
\begin{align}\label{F factor eta0=etaA}
    F\sim&2\pi d\epsilon\frac{1}{3\eta_I^A(1+w)(3\eta_I^A(1+w)+d(1+3w))}\nonumber\\&-16\epsilon^2\ln\left(\frac{d}{\epsilon}\right)\frac{1}{(3\eta_I^A(1+w)+d(1+3w))^2}\,.
\end{align}\normalsize

\begin{figure*}
    \centering
    \includegraphics[scale=1]{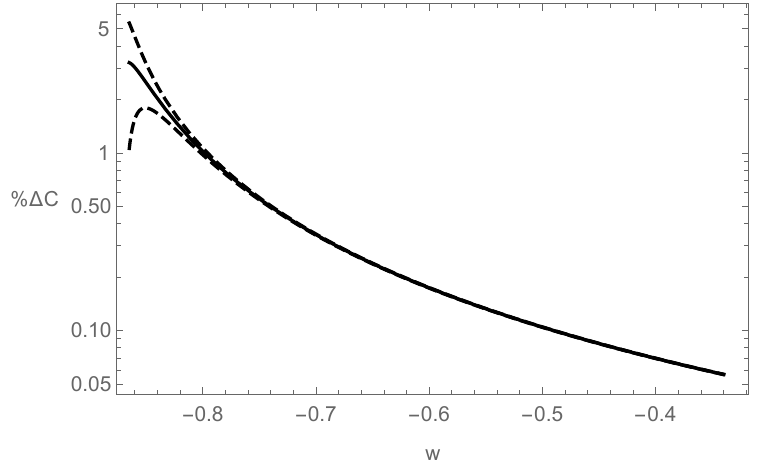}
    \caption{Plot of the increasing percentage of the classical capacity $\%\Delta C$ from Eq.~\eqref{relative increasing of the capacity} in case of minimal coupling $\xi=0$, for different values of the barotropic index $w$. The other parameters are $\epsilon=0.01$, $\eta_I^A=5$ and $d=1$. The dashed lines represent the minimum and maximum values of $\%\Delta C$ by taking into account the maximum relative error $\mathcal{E}$ in Eq.~\eqref{Maximum relative error}.}
    \label{fig:Capacity increasing for w}
\end{figure*}

A plot of $F$, corresponding to $\Delta C$ in case of minimal coupling $\xi=0$ (see Eq.~\eqref{relative increasing of the capacity}), is shown in terms of $w$ in Fig.~\ref{fig:Capacity increasing for w}. The maximum relative error, $\mathcal{E}$, in $F$ is taken into account with the dashed lines, bounding the possible values of $F$, including this error.

From Fig.~\ref{fig:Capacity increasing for w}, we display different increasing of the classical capacity - denoted with $\%\Delta C=100\cdot\Delta C/C_M$ - for different values of $w$. This shows that the Wi-Fi communication protocol between quantum devices can be exploited to achieve information about the current cosmological expansion - in this case, parameterized only by $w$.

Moreover, as specified in Eq.~\eqref{relative increasing of the capacity}, one can also get information about the coupling $\xi$ between the field and the curvature. In particular, one could get an increase of the capacity which is lower than the one predicted in Fig.~\ref{fig:Capacity increasing for w}, meaning that the coupling between quantum field and scalar curvature would be different than the minimal one.

It is worth noticing, from Fig.~\ref{fig:Capacity increasing for w} and Eq.~\eqref{F factor eta0=etaA}, that $\Delta C/C_M|_{\xi=0}=F$ increases by decreasing $w$ to reach a divergence at a certain point.

This is because, at $w=-\frac{d+3\eta_I^A}{3(d+\eta_I^A)}$, the detector $B$ is positioned at the sender's comoving horizon $d_H$ at the time $\eta_I^A$, reading
\begin{equation}\label{Alice's comoving horizon}
    d_H=-3\eta_I^A\left(\frac{1+w}{1+3w}\right)\,.
\end{equation}
When $d>d_H$, the signal sent by the detector $A$ would never reach the detector $B$. Then, it is worth studying the factor $F$ with the scaled quantities
\begin{equation}\label{scaling}
    \tilde{\epsilon}\coloneqq\frac{\epsilon}{d_H}\,;\quad\tilde{d}\coloneqq\frac{d}{d_H}\,,
\end{equation}
where $0<\tilde{\epsilon}<1$, $0<\tilde{d}<1$ and $\tilde{\epsilon}\ll\Tilde{d}$.

With them, the factor $F$ from Eq.~\eqref{F factor eta0=etaA} becomes
\begin{equation}\label{F factor normalized}
    F\sim\frac{2\pi}{(1+3w)^2}\frac{\tilde{\epsilon}\tilde{d}}{1-\tilde{d}}-\frac{16}{(1+3w)^2}\frac{\tilde{\epsilon}^2}{(1-\tilde{d})^2}\ln\left(\frac{\tilde{d}}{\tilde{\epsilon}}\right)\,.
\end{equation}
Moreover, by using $\tilde{\epsilon}$ and $\tilde{d}$ a simple analytical expression could by found also for $\mathcal{E}_P$ and $\mathcal{E}_F$ - from Eqs.~\eqref{Perturbation error main text} and \eqref{Fermi bound error}, respectively - giving the relative error of $F$ through Eq.~\eqref{Maximum relative error}, namely
\begin{equation}\label{perturbation error exp}
    \mathcal{E}_P=32\frac{6\tilde{d}+2(3-\tilde{d})\ln\left(1-\tilde{d}\right)-\ln^2\left(1-\tilde{d}\right)}{(1+3w)^4}\,.
\end{equation}
\begin{equation}\label{Fermi bound approximation normalized}
    \mathcal{E}_F=-\frac{\Tilde{\epsilon}}{(1+3w)(1-\tilde{d})}\,;
\end{equation}
From Eq.~\eqref{F factor normalized}, we see that the divergence occurring in Eq.~\eqref{F factor eta0=etaA} when $w$ gets close to $-1$ disappears. Moreover, from Eqs.~\eqref{Fermi bound approximation normalized} and \eqref{perturbation error exp}, also the error do not increase indefinitely in this limit. From Eq.~\eqref{Alice's comoving horizon}, one may argue that, in the limit $w\to-1$, the horizon gets too small to allow a physically reasonable setup for the communication protocol. However, from Eq.~\eqref{Alice's comoving horizon}, as long as $w\ne-1$, one could always consider $\eta_I^A$ high enough to allow any size for the detectors $\epsilon$ and their mutual distance $d$.
    \begin{figure*}
\centering
    \includegraphics[scale=1]{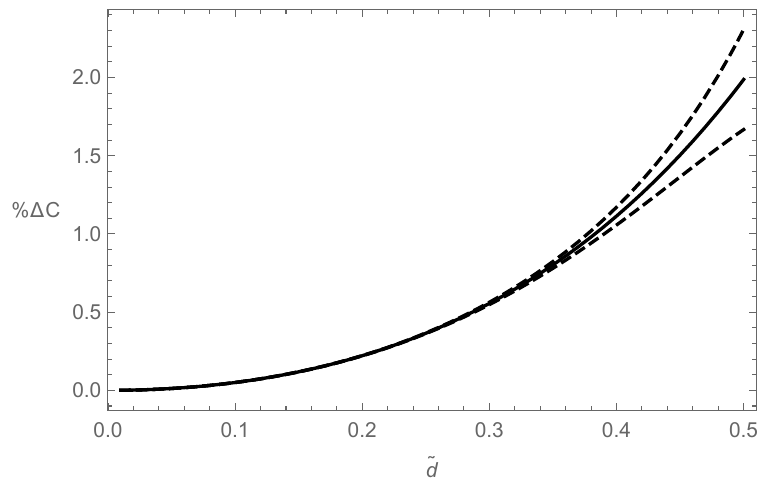}
    \caption{Plot of the percentage increase of the classical capacity with respect to the Minkowski case when $\xi=0$ (minimal coupling) vs the normalized distance $\tilde{d}$. The parameters are fixed to be $\tilde{\epsilon}=0.01\cdot d$ and $w=-0.7$. The dashed lines indicate the maximum and minimum value that $\%\Delta C$ can assume by including the error $\mathcal{E}$ given by Eq.~\eqref{Maximum relative error}.}
    \label{fig:CapacityIncreasingRunD}
\end{figure*}

In Fig.~\ref{fig:CapacityIncreasingRunD} the relative increase of the classical capacity $\Delta C$ - considering the minimal coupling $\xi=0$, so that $\Delta C=F$ - is shown for different values $\tilde{d}$ and keeping fixed the ratio $\epsilon/d$. The dashed lines represent the minimum and maximum values of $F$ by considering the relative error $\mathcal{E}$. From Fig.~\ref{fig:CapacityIncreasingRunD} we see that the relative increase of the capacity is higher the more $\epsilon$ and $d$ enlarge by keeping fixed their ratio. In other words, $\Delta C$ is higher the more we consider the two-detectors system to be "large". As the detector $B$ approaches the sender's comoving horizon, $\mathcal{E}$ diverges and then both the Fermi bound approximation and the first order perturbation theory are no longer valid. Nevertheless, the trend given in the range when $\mathcal{E}\ll1$ suggests interesting communication properties achievable as $\tilde{d}$ grows to become closer to $1$.

\subsection{The case of de Sitter universe}

The specific case, $w=-1$, represents a very relevant cosmological scenario, typically associated with strongly-accelerated phases of the universe \cite{Dunsby:2015ers}. At small redshifts, the standard background model, i.e., the $\Lambda$CDM paradigm, is exactly based on the existence of a (bare) cosmological constant. Its validity is limited to intermediate redshifts, up to which dark energy seems to be described by it successfully\footnote{Naive extensions of the standard cosmological background could even be  used to heal the cosmological constant problem \cite{Luongo:2018lgy,Belfiglio:2022qai} and are currently under debate in view of the recent developments offered by the DESI collaboration, see e.g. \cite{DESI:2024mwx}.} \cite{Luongo:2020hyk}.  On the other side, at primordial times inflation represents a phase of de Sitter expansion \cite{Ellis:2023wic}, compatible with $w=-1$.

Recent cosmological tensions \cite{Hu:2023jqc}, cosmological observations of possibly evolving dark energy \cite{Luongo:2024fww,Carloni:2024zpl} and issues related to the nature of the inflaton \cite{Luongo:2024opv} leave open the possibility that, rather than a genuine de Sitter phase, the universe can be characterized by a quasi-de Sitter epoch, in which $w\simeq-1$ \cite{Geshnizjani:2023hyd}. We start with the latter, as it turns out to be quite relevant in inflationary particle production \cite{Belfiglio:2022cnd,Belfiglio:2022yvs,Belfiglio:2023moe,Belfiglio:2024xqt}.

Particularly, as in the EdS case, Eq.~\eqref{Friedmann Eq. perfect fluid} is solved in cosmic time, $t$, giving a scale factor
\begin{equation}\label{w0=-1 scale}
    a(t)=e^{H_0(t-t_I^A)}\,,
\end{equation}
where $H_0\coloneqq\dot{a}(t_I^A)/a(t_I^A)$.

The relation between the conformal time $\eta$ and the cosmic time $t$, by imposing $a(\eta_I^A)=1$, is given by
\begin{equation}
    \eta-\eta_I^A=\int_{t_I^A}^t e^{-H_0(t'-t_I^A)}dt'=-\frac{1}{H_0}(e^{-H_0(t-t_I^A)}-1)\,,
\end{equation}
where $t_I^A$ is taken into account, instead of $-\infty$.

In this naive picture, the scale factor acquires the form
\begin{equation}\label{w=-1 scale factor conf time}
    a(\eta)=\frac{1}{1+H_0(\eta_I^A-\eta)}\,,
\end{equation}
finally ending up with
\begin{equation}\label{solution de Sitter}
    aH_c(\eta)=\frac{H_0}{(1+H_0(\eta_I^A-\eta))}\,.
\end{equation}
Adopting the same strategy above discussed, plugging  Eq.~\eqref{solution de Sitter} into Eq.~\eqref{factor F main text}, one obtains an exact solution for $F$.

Hence, following Eq.~\eqref{F factor eta0=etaA} and bearing the same assumptions used for the EdS case, one obtains
\begin{equation}\label{F factor dS}
    F\sim\frac{\epsilon H_0^2}{2}\left(\frac{\pi d}{1-dH_0}-\frac{8\epsilon}{(1-dH_0)^2}\ln\left(\frac{d}{\epsilon}\right)\right)\,.
\end{equation}
From Eq.~\eqref{w=-1 scale factor conf time}, we immediately see that the sender's comoving horizon is here $d_H=H_0^{-1}$.

Thus, by using $\tilde{\epsilon}$ and $\tilde{d}$ as defined in Eq.~\eqref{scaling}, Eq.~\eqref{F factor dS} yields
\begin{equation}\label{F factor dS normalized}
    F\sim \frac{\pi}{2}\frac{\tilde{\epsilon}\tilde{d}}{1-\tilde{d}}-\frac{4\tilde{\epsilon}^2}{(1-\tilde{d})^2}\ln(\tilde{d}/\tilde{\epsilon}).
\end{equation}
The errors given by the first order perturbation theory and by the Fermi bound, from Eqs.~\eqref{Perturbation error main text} and \eqref{Fermi bound error} respectively, are
\begin{align}
    \mathcal{E}_P&=2\left(6\tilde{d}+2(3-\tilde{d})\ln(1-\tilde{d})-\ln^2(1-\tilde{d})\right)\,,\label{perturbation error dS}\\
    \mathcal{E}_F&=\frac{\tilde{\epsilon}}{2(1-\tilde{d})}\,.\label{Fermi bound error dS}
\end{align}
We can see that Eqs.~\eqref{F factor dS normalized}, \eqref{perturbation error dS} and \eqref{Fermi bound error dS} are identical respectively to Eqs.~\eqref{F factor normalized}, \eqref{perturbation error exp} and \eqref{Fermi bound approximation normalized} in the limit $w\to-1$.

The de Sitter expansion is recovered when $H_0=-\frac{1}{\eta_I^A}$. In so doing, the scale factor in Eq.~\eqref{w=-1 scale factor conf time} becomes
\begin{equation}
    a(\eta)=-\frac{1}{H_0\eta}\,,
\end{equation}
where $\eta<0$. In the de Sitter case, also Eqs.~\eqref{solution de Sitter} and \eqref{F factor dS} are valid as long as $H_0=-\frac{1}{\eta_I^A}$. The sender's comiving horizon becomes then $d_H=-\eta_I^A$. Then, by using the scaling \eqref{scaling}, the $F$ factor in a de Sitter expansion, where $w=-1$, corresponds to the one in Eq.~\eqref{F factor dS}, with an estimated relative error given by Eqs.~\eqref{Maximum relative error}, \eqref{perturbation error dS} and \eqref{Fermi bound error dS}.

\begin{figure*}
    \centering
    \includegraphics[scale=1]{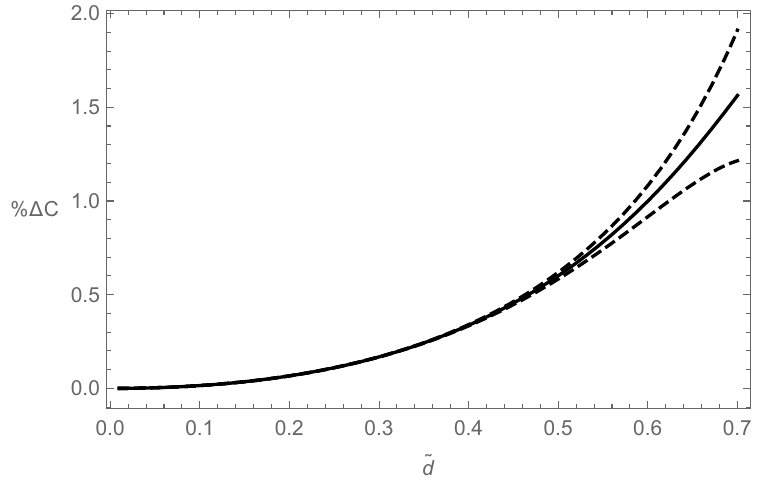}
    \caption{Plot of the percentage increase of the classical capacity with respect to the Minkowski case when $\xi=0$ (minimal coupling) vs the normalized distance $\tilde{d}$. The parameters are fixed to be $\tilde{\epsilon}=0.01\cdot d$ and $w=-1$. The dashed lines indicate the maximum and minimum value that $\%\Delta C$ can assume by including the error $\mathcal{E}$ given by Eq.~\eqref{Maximum relative error}.}
    \label{fig:capacity increasing run d dS}
\end{figure*}
The behaviour of $\Delta C$, in the minimal coupling case $\xi=0$ and when $w=-1$, is plotted in Fig.~\ref{fig:capacity increasing run d dS}. By comparing the case $w=-0.7$, in Fig.~\ref{fig:CapacityIncreasingRunD}, with the case $w=-1$ in Fig.~\ref{fig:capacity increasing run d dS}, we see that the increment of the classical capacity is lower the closer $w$ is to $-1$. In fact, from Eq.~\eqref{F factor normalized} we see that $F=\Delta C$ scales as $(1+3w)^{-2}$. However, from Figs.~\ref{fig:CapacityIncreasingRunD} and \ref{fig:capacity increasing run d dS} we also notice that the error is lower in the case $w=-1$. This is because the error $\mathcal{E}_P$ in Eq.~\eqref{perturbation error exp} - expected to dominate over $\mathcal{E}_F$ when $1-\tilde{d}\gg\tilde{\epsilon}$ - scales as $(1+3w)^{-4}$.

As a consequence, despite the increasing of the capacity is lower. In the case $w=-1$ we can better explore the transmission of information when the receiver is closer to the sender's horizon.

\section{Outlooks and perspectives}

The transmission of classical information between two harmonic oscillator detectors has been examinated while considering an expanding universe background. The method used consists on the Heisenberg evolution of the harmonic oscillators as they interact with the field via a rapid interaction. This enabled the derivation of a bosonic one-mode Gaussian channel for the communication protocol, whose properties and capacities are possible to find non-perturbatively.

Specifically, by constraining the energy of the detectors, the classical capacity of the channel turned out to be similarly constrained. This bound depends solely on the properties of the background spacetime through which the quantum field propagates. Consequently, various gravitational fields can be discrimated by examining the maximum rate at which the channel can reliably transmit classical information.

The impact of cosmological expansion on the transmission of classical information has been explored. In particular, we wondered whether the universe expansion might enhance the communication capabilities of the detectors with respect to a flat spacetime scenario - or if it introduces an additional obstacle to the communication. Our findings indicate that the effect of cosmological expansion on the channel is primarily dependent on the coupling between the field and the scalar curvature. Specifically, a conformally coupled field $\xi=1/6$ does not influence the classical capacity of the channel. Conversely, in the case of minimal coupling $\xi=0$, except for exceptional situations - such as a universe that expands abruptly just after the sender's interaction with the field - a cosmological expansion generally enhances the channel's ability to transmit classical messages.

By considering an example of cosmological expansion driven by a perfect fluid, we consistently observe an increase in the classical capacity. This increase is dependent on the barotropic index $w$ of the fluid. Furthermore, we found that the increase in capacity is larger when the detectors are bigger and farther apart.

Consequently, to observe a significant increase in capacity, the physical system might be as large as possible. Naturally, if the oscillator of the receiver is close to the comoving horizon of the sender, the two detectors can no longer be considered non-relativistic.

Future research will address relativistic particle detector models to investigate general communication properties near horizons. As an alternative, one can also consider point-like detectors and drop the hypothesis of a rapid interaction \cite{Louko_2006}.

In addition, detecting the quantum effects here studied would require highly precise instruments or very large experimental setups - similar to those needed to detect the Unruh effect via particle detectors \cite{D_Bunney_2023}. However, recent advancements in analog models of cosmological expansion, which predict observable quantum effects, have been developed in laboratory settings \cite{Tian_2017,Steinhauer_2022}. Thus, our communication protocol could be more feasibly implemented in a laboratory using analog gravity systems \cite{Jacquet_2020}, offering the potential to observe the coupling between the communicated particles and the emulated curvature.

Furthermore, future research will focus on the potential for reliable communication of quantum messages in an expanding cosmological background. According to the no-cloning theorem, reliable quantum communication is not possible in isotropic systems \cite{Jonsson_2018,lapponi2024making}. To address this issue, one could investigate whether an anisotropic cosmological expansion might enable a quantum capacity larger than zero.

\section*{Acknowledgements}

A.L. acknowledges his stay at the University of Nottingham, UK, during which the present work has been conceived. He also acknowledges Salvatore Capozziello for crucial discussions to obtain the shown results. O.L. acknowledges Rocco D'Agostino, Francesco Pace and Sunny Vagnozzi for interesting discussions. S.M. acknowledges support from Italian Ministry of Universities and Research under “PNRR MUR project PE0000023-NQSTI”.

\bibliographystyle{unsrt}

\newpage
\onecolumngrid
\appendix

\section{Fermi Bound}\label{appendix: Fermi Bound}

To consider a non-relativistic detector in a curved background, we need the proper coordinates of the detectors to be non-relativistic at least locally. A set of locally non-relativistic coordinates for the detector $i$ is given by the \textit{Fermi-normal coordinates} $(t_i,\mathbf{x}_i)$, where $t_i$ is the proper time of the detector $i$ and $\mathbf{x}_i=(x_i,y_i,z_i)$ are spatial coordinates required to be orthonormal to the four-velocity of the detector at each time $i$. The size of the region around the detector $i$, where the non-relativistic coordinates $(t_i,\mathbf{x}_i)$ can be used is called \textit{Fermi length} $L_F$. In Ref.~\cite{Perche_2022}, this length was estimated to be
\begin{equation}\label{Fermi bound general}
L_F^i\sim\frac{1}{\alpha_i+\sqrt{\lambda_R^i}}\,,
\end{equation}
where $\alpha_i$ is the proper acceleration of the detector and $\lambda_R^i$ is the greatest eigenvalue of the Riemann tensor component $R_{0j0k}$ computed in the detector $i$'s center of mass at the time $t_i$ ($j,k$ indicate spatial coordinates).

For static detectors in a background given by the FRLW metric, Eq.~\eqref{FLRW metric}, we have $t_i=t$ and, then, since the detectors are positioned as indicated in Eq.~\eqref{detectors' position}, the relation between the coordinates $\mathbf{x}$ and $\mathbf{x}_i$ read
\begin{equation}\label{Fermi normal coordinates}
    \begin{cases}
        &\mathbf{x}_A=a(t)\left(\mathbf{x}-\mathbf{d}\right)\,,\\
        &\mathbf{x}_B=a(t)\mathbf{x}\,.
    \end{cases}
\end{equation}
By using the coordinates \eqref{Fermi normal coordinates}, one can easily compute $\lambda_R$. Since $\alpha_i=0$, we have the Fermi length
\begin{equation}\label{Fermi bound CE appendix}
    L_F=\frac{a}{\sqrt{\dot{a}^2-\ddot{a}a}}=\sqrt{\frac{6}{R}}\,.
\end{equation}

\section{Classical capacity of a one-mode Gaussian channels}\label{appendix: capacities}

In this Appendix we recall the classical capacity of a one-mode Gaussian quantum channel with input energy constraint.

A message corresponds to n-realizations of a random variable $X$, i.e. to e sequence $x^n\equiv x_1\ldots x_n$ picked up with probability $p_{x^n}$. The sequence is then encoded into an $n$-mode state $\rho_{x^n}$. Upon sending it through $n$ uses of the channel ${\cal N}$, the receiver will get ${\cal N}^{\otimes n}(\rho_{x^n})$ and from it decode a sequence $y^n\equiv y_1\ldots y_n$ constituting n realizations of a random variable $Y$.

The information about $X^n$ present in $Y^n$ depends on the encoding and decoding operations. However it is known that the mutual information between random variables $X^n$ and $Y^n$ is upper bound by the \emph{Holevo information} \cite{Holevo1973BoundsFT}
\begin{equation}
\chi\left(\rho^{(n)},{\cal N}^{\otimes n}\right)=S\left(\sum_{x^n} p_{x^n} {\cal N}^{\otimes n}(\rho_{x^n})\right)-S\left({\cal N}(\rho^{(n)})\right),
\end{equation}
where  $\rho^{(n)}=\sum_{x^n}p_{x^n}\rho_{x^n}$ is the average input state.

As a consequence, the classical capacity of the channel ${\cal N}$ becomes
\begin{equation}
C({\cal N})=\lim_{n\to\infty}\frac{1}{n}\max_{\rho^{(n)}} \chi\left(\rho^{(n)},{\cal N}^{\otimes n}\right).
\end{equation}

If the encoding is done into product states of the kind $\rho_{x^n}=\otimes_{i=1}^n\rho_{x_i}$, then, the capacity formula reduces to
a single letter version
\begin{equation}\label{eq:C1}
C^{(1)}({\cal N})=\max_{\rho} \chi\left(\rho,{\cal N}\right)
\end{equation}
where
\begin{equation}\label{Holevo bound}
\chi(\rho,{\cal N})=S\left(\sum_x p_x {\cal N}(\rho_x)\right)-S\left({\cal N}(\rho)\right)
\end{equation}
and $\rho=\sum_xp_x\rho_x$.

Eq.\eqref{eq:C1} is known as \emph{product state capacity} and it is always $C^{(1)}({\cal N})\leq C({\cal N})$. When the equality holds, the channel $\mathcal{N}$ is called \textit{additive}. This is e.g. the case of \textit{entanglement-breaking channels} \cite{Shor_2002}, mentioned in Sec.~\ref{sec communication protocol}. In Sec.~\ref{sec: Cosmological expansion} we see that the channel we consider throughout the paper is entanglement-breaking, so that we can write
\begin{equation}\label{Classical capacity from HI}
    C(\mathcal{N})=C^{(1)}(\mathcal{N})=\max_{\rho}\mathcal{X}(\rho,\mathcal{N})\,.
\end{equation}

To compute the classical capacity from Eq.~\eqref{Classical capacity from HI}, one needs to maximize Eq.~\eqref{Holevo bound} over all the bosonic states $\rho$. However, it is conjectured that bosonic Gaussian states are the ones preserving better the encoded classical information \cite{giovannetti2015solution}. Henceforth, we exclusively consider Gaussian input states. Then, by considering a one-mode Gaussian state mapping as in Eq.~\eqref{input-output relation} the Holevo information results \cite{Lupo_2011,Pilyavets_2012}
\begin{equation}\label{Holevo information one-mode Gaussian channels}
    \mathcal{X}(\sigma_{in},\sigma_{enc},\mathcal{N})=\mathcal{S}\left(\mathbb{T}(\sigma_{in}+\sigma_{enc})\mathbb{T}^T+\mathbb{N}\right)-\mathcal{S}\left(\mathbb{T}\sigma_{in}\mathbb{T}^T+\mathbb{N}\right)\,,
\end{equation}\normalsize
where we suppose that the sender encodes the classical message into $\sigma_{in}+\sigma_{enc}$. The classical capacity of a one-mode Gaussian channel thus reads
\begin{equation}
    C(\mathcal{N})=\max_{\sigma_{in},\sigma_{enc}}\mathcal{X}(\sigma_{in},\sigma_{enc},\mathcal{N})\,.
\end{equation}
By considering the generic form of a one-mode covariance matrix \eqref{generic detector state}, we can write
\begin{equation}\label{input covariance matrix}
    \sigma_{in}\equiv\text{diag}\left(\left(\frac{1}{2}+N_{in}\right)e^{l_{in}},\left(\frac{1}{2}+N_{in}\right)e^{-l_{in}}\right)\,,
\end{equation}
\begin{equation}\label{encoding covariance matrix}
    \sigma_{in}+\sigma_{enc}\equiv\text{diag}\left(\left(\frac{1}{2}+N_{enc}\right)e^{l_{enc}},\left(\frac{1}{2}+N_{enc}\right)e^{-l_{enc}}\right)\,.
\end{equation}\normalsize

Focusing on the Holevo information \eqref{Holevo information one-mode Gaussian channels}, it is clear that $\mathcal{X}$ increases arbitrarily by increasing $N_{enc}$ while keeping $N_{in}$ finite.

In this way, the classical capacity of a one-mode Gaussian channel is always infinite unless we impose an upper bound for $N_{enc}$. This upper bound would be reasonable, since an unbounded $N_{enc}$ means that the sender can encode an arbitrarily high amount of information in her detector.

For the particle detectors, described in Sec.~\ref{sec: oscillator hamiltonian}, an upper bound for $N_{enc}$ is naturally provided by the energetic condition \eqref{energetic condition for the detector}, which becomes, together with Eq.~\eqref{detector energy},
\begin{equation}\label{Encoding N condition}
    N_{enc}\le\frac{E_c^A}{\omega_A\cosh l_{enc}}-\frac{1}{2}\,.
\end{equation}

The Holevo information, Eq. \eqref{Holevo information one-mode Gaussian channels}, requires now to be maximized over $\sigma_{in}$ and $\sigma_{enc}$ while respecting the bound \eqref{Encoding N condition}.

This problem was faced in Ref.~\cite{Lapponi_2023} where, by considering the parameters $\tau$ and $W$, corresponding to the determinant of $\mathbb{T}$ and $\mathbb{N}$, respectively, the Holevo information \eqref{Holevo information one-mode Gaussian channels} was optimized apart from a parameter $J$ bounded from $0$ to $2\frac{E_c^A}{\omega_A}+\sqrt{4\left(\frac{E_c^A}{\omega_A}\right)^2-1}$. Hence,
\begin{align}
    \mathcal{X}(J)=h\left(\frac{1}{2}\sqrt{\left(\tau J+2\sqrt{W}+2\tau X\right)\left(4\tau\frac{E_c^A}{\omega_A}+2\sqrt{W}-\tau J-2\tau X\right)}\right)-h\left(\frac{1}{2}\sqrt{\left(\tau J+2\sqrt{W}\right)\left(\frac{\tau}{J}+2\sqrt{W}\right)}\right)\,,\label{Holevo information for J}
\end{align}\normalsize
where $
    h(x)\equiv \left(x+\frac{1}{2}\right)\log\left(x+\frac{1}{2}\right)-\left(x-\frac{1}{2}\right)\log\left(x-\frac{1}{2}\right)$,
and
\begin{align}\label{xmax}
    X=\left\{\begin{matrix*}
    0, &\text{if}&  2\frac{E_c^A}{\omega_\textrm{A}}<J<2\frac{E_c^A}{\omega_\textrm{A}}+\sqrt{4\frac{(E_c^A)^2}{\omega_\textrm{A}^2}-1}\,,\\
        \frac{E_c^A}{\omega_\textrm{A}}-\frac{J}{2}, &\text{if}& \frac{\omega_\textrm{A}}{2E_c^A}<J<2\frac{E_c^A}{\omega_\textrm{A}}.
        \end{matrix*}
        \right.
\end{align}\normalsize

Thus, to find the classical capacity, one has essentially to maximize Eq.~\eqref{Holevo information for J} for $J$. It can be proved that, for each $\tau$ and $W$, we have $\partial_J\mathcal{X}(J=1)=0$ and $\partial_J^2\mathcal{X}(J=1)<0$. Then, the maximum of $\mathcal{X}$ from Eq.~\eqref{Holevo information for J} is provided when $J=1$. The constrained classical capacity for our channel is therefore
\begin{equation}\label{Classical capacity of the channel appendix}
    C(\tau,W)=h\left(\frac{E_c^A}{\omega_A}\tau+\sqrt{W}\right)-h\left(\frac{\tau}{2}+\sqrt{W}\right)\,.
\end{equation}

\section{Calculations of the transmissivity in a perturbative cosmological expansion}\label{appendix: perturbation}

In this appendix, the transmissivity in Eq.~\eqref{maximum transmissivity 3} explicitly computed. The Lagrangian density of a massless scalar field, coupled to the scalar curvature $R$ in a FLRW background reads \cite{NDBirrell_1980,Ford1987}
\begin{equation}\label{lagrangian density}
    \mathcal{L}=\frac{1}{2}g^{\mu\nu}\partial_\mu\Phi\partial_\nu\Phi+\frac{1}{2}\xi R\Phi^2\,,
\end{equation}
where $\xi$ is the curvature-field coupling, giving to the field a time dependent effective mass $\sqrt{\xi R}$.

By choosing the Minkowski vacuum $\ket{0}$ as reference vacuum, the field operator can be expanded as
\begin{equation}\label{field operator expansion}
    \hat{\Phi}(\mathbf{x},t)=\int d\mathbf{k}\left(a_\mathbf{k}u_\mathbf{k}(\mathbf{x},t)+a^\dagger_{\mathbf{k}}u^\ast_{\mathbf{k}}(\mathbf{x},t)\right)\,,
\end{equation}
where $a_{\mathbf{k}}\ket{0}=0$. The modes $u_\mathbf{k}(\mathbf{x},t)$ are the solutions of the Klein-Gordon equation - obtained from the Lagrangian \eqref{lagrangian density}
\begin{equation}\label{generalized KG equation}
    \left(\Box+\xi R\right)u_\mathbf{k}=0\,,
\end{equation}
where $\Box$ is the D'Alembert operator. Following the normalization condition\small
\begin{equation}
    \left(u_{\mathbf{k}},u_{\mathbf{k}'}\right)=-i\int_{\Sigma_t}\left(u_{\mathbf{k}}\partial_t u_{\mathbf{k}'}^\ast+u_{\mathbf{k}'}\partial_tu_{\mathbf{k}}^\ast\right)\sqrt{-g}d\mathbf{x}=\delta^3(\mathbf{k}-\mathbf{k}')\,,
\end{equation}\normalsize
where $\Sigma_t$ is the Cauchy surface $t=\text{const}$ and, considering Eq.~\eqref{FLRW metric}, the solutions of Eq.~\eqref{generalized KG equation} can be written in the form \cite{Ford1987}
\begin{equation}\label{normal mode decomposition}
    u_\mathbf{k}(\mathbf{x},t)=\frac{e^{i\mathbf{k}\cdot\mathbf{x}}}{a(\eta)\sqrt{(2\pi)^3}}\chi_k(\eta(t))\,,
\end{equation}
where $k\equiv|\mathbf{k}|$, $\eta$ is the conformal time, i.e., satisfying $a(\eta)d\eta=dt$, and $\chi_k$ leads to
\begin{equation}\label{equation for chi}
    \chi_k''(\eta)+(k^2+U(\eta))\chi_k(\eta)=0\,,
\end{equation}
where the prime $'$ denotes a derivative with respect to $\eta$. The corresponding potential in a homogeneous and isotropic universe acquires the form,
\begin{equation}\label{potential appendix}
    U(\eta)\coloneqq \left(6\xi-1\right)\left(\frac{a'(\eta)}{a(\eta)}\right)^2=(6\xi-1)a^2(\eta)H_c^2(\eta)\,,
\end{equation}
where $H_c\equiv a'(\eta)/a^2(\eta)$, usually dubbed \textit{conformal Hubble parameter}.

Since the spacetime curvature is considered negligible at $t=t_I^A$, in a neighborhood of $t_I^A$ the modes \eqref{normal mode decomposition} read\\
\begin{equation}\label{boundary condition for the mode}
    u_\mathbf{k}(\mathbf{x},t\sim t_I^A)=\frac{e^{i\mathbf{k}\cdot\mathbf{x}-ikt}}{\sqrt{2k}(2\pi)^{3/2}a(t_I^A)}\,.
\end{equation}
Given that $a(t_I^A)$ has only the role of scaling the initial size of the detectors and their distance, we set $a(t_I^A)=1$ for the sake of simplicity.

An analytical solution of $\chi_k(\eta)$ is achievable exclusively in some particular cases \cite{Ford_2021}. Hence, to reach a viable solution, we may use a perturbation method \cite{Zeldovich:1971mw}, based on defining $\eta_I^A=\eta(t_I^A)$ and rewriting Eq.~\eqref{equation for chi} in an integral form:
\begin{equation}\label{integral form of the chi equation}
    \chi_k(\eta)=\chi_k^{(0)}(\eta)-\frac{1}{k}\int_{\eta_I^A}^\eta\sin(k(\eta-\eta_1))U(\eta_1)\chi_k(\eta_1)d\eta_1\,,
\end{equation}\normalsize
where $\chi_k^{(0)}$ corresponds to the expression of $\chi_k(\eta)$ acquired when $U(\eta)=0$, i.e., from Eq.~\eqref{boundary condition for the mode}
\begin{equation}\label{conformal coupling solution}
    \chi_k^{(0)}(\eta)=\frac{e^{-ik\eta}}{\sqrt{2k}}\,.
\end{equation}
By defining the integral operator $\hat{\mathcal{U}}$, acting on a test function $f(t)$ as
\begin{equation}
    \hat{\mathcal{U}}\left[f(\eta)\right]=-\frac{1}{k}\int_{\eta_I^A}^\eta\sin(k(\eta-\eta_1))U(\eta_1)f(\eta_1)\,,
\end{equation}
immediately Eq.~\eqref{integral form of the chi equation} can be rewritten as
\begin{equation}\label{equation for chi operatorial form}
    \chi_k^{(0)}(\eta)=\left(\text{Id}-\hat{\mathcal{U}}\right)\chi_k(\eta)\,.
\end{equation}
Thus, Eq.~\eqref{equation for chi operatorial form} can be then inverted to obtain
\begin{equation}\label{complete solution sum}
    \chi_k(\eta)=(\text{Id}-\hat{\mathcal{U}})^{-1}\chi_k^{(0)}(\eta)=\left(\sum_{l=0}^\infty\hat{\mathcal{U}}^l\right)\chi_k^{(0)}(\eta)\,.
\end{equation}
For example, by making explicit only the first three terms of the sum at the r.h.s. of Eq.~\eqref{complete solution sum} we have
\footnotesize

\begin{align}
    \chi_k&(\eta)=\chi^{(0)}_k(\eta)-\frac{1}{k}\int_{\eta_I^A}^\eta\sin(k(\eta-\eta_1))U(\eta_1)\chi_k^{(0)}(\eta_1)d\eta_1+\frac{1}{k^2}\int_{\eta_I^A}^{\eta}\int_{\eta_I^A}^{\eta_1}\sin(k(\eta-\eta_1))\sin(k(\eta_1-\eta_2)) U(\eta_1)U(\eta_2)\chi_k^{(0)}(\eta_2)d\eta_1d\eta_2+\mathcal{O}(U(\eta)^3)\,.
\end{align}
\normalsize
Clearly our approximation lies on assuming that $U(\eta)$ is small enough within the interval [$\eta_I^A$,$\eta$], since in this case only the first terms of Eq. ~\eqref{complete solution sum} can be considered, neglecting higher order ones.

In other words, if $U(\eta)$ is small enough, the solutions $\chi_k$ can be expressed as $\chi^{(0)}(\eta)$ plus some perturbative terms.

We consider the first order perturbation theory, leading to the approximation
\begin{align}
    \chi_k(\eta)\sim (\text{Id}+\hat{\mathcal{U}})\chi^{(0)}_k(\eta)=\frac{e^{-ik\eta}}{\sqrt{2k}}+\frac{i}{(2k)^{3/2}}\left(6\xi-1\right)\int_{\eta_I^A}^{\eta}\left(e^{ik(\eta-2\eta_1)}-e^{-ik\eta}\right)a^2(\eta_1)H_c^2(\eta_1)d\eta_1\,.\label{chi perturbative}
\end{align}

The relative error we carry over by neglecting further perturbative terms can be estimated by the ratio $\frac{\hat{\mathcal{U}}^2\chi_k^{(0)}(\eta)}{\chi_k^{(0)}(\eta)}$, becoming maximum in the limit $k\to0$, leading to the maximum relative error, $\mathcal{E}_P$,
\begin{align}
    \lim_{k\to0}\frac{\hat{\mathcal{U}}^2\chi_k^{(0)}(\eta)}{\chi_k^{(0)}(\eta)}=\int_{\eta_I^A}^\eta\int_{\eta_I^A}^{\eta_1}(\eta-\eta_1)(\eta_1-\eta_2)U(\eta_1)U(\eta_2)d\eta_1d\eta_2\,.\label{Perturbation error chi}
\end{align}
At this point, we can compute the integral $I(t_I^A,t_I^B)$ from Eq.~\eqref{I integral def}. Using the Lorentzian smearing, Eq. \eqref{Lorentzian shape}, and the mode decomposition, Eq. \eqref{normal mode decomposition}, then the integrals over $\mathbf{x}$ and $\mathbf{x}'$ give
\begin{align}
    I(\eta_I^A,\eta_I^B)=\frac{2}{(2\pi)^2da(\eta_I^A)a(\eta_I^B)}\Re\int_0^\infty k\chi_k(\eta_I^A)\chi_k^\ast(\eta_I^B)\left(e^{ikd-2\epsilon k}-e^{-ikd-2\epsilon k}\right)dk\,.\label{I integral cosmological expansion pass 1}
\end{align}
The  product between the functions $\chi_k$ in the above integral, up to first order, reads from Eq.~\eqref{chi perturbative}
\begin{align}
    \chi_k(\eta_I^A)\chi^\ast_k(\eta_I^B)=&\frac{e^{-ik(\eta_I^A-\eta_I^B)}}{2k}-\frac{i}{(2k)^2}\left(6\xi-1\right)\left(-e^{-ik(\eta_I^A-\eta_I^B)}\int_{\eta_I^A}^{\eta_I^B}a^2(\eta)H_c^2(\eta)d\eta+\int_{\eta_I^A}^{\eta_I^B}e^{-ik(\eta_I^A+\eta_I^B-2\eta)}a^2(\eta)H_c^2(\eta)d\eta\right)\,.\label{chi product}
\end{align}\normalsize

Since $d$ is the conformal distance between the detectors, from the Huygens principle we expect the higher transmissivity to occur when $\eta_I^B-\eta_I^A=d$. Accordingly, we exclusively focus on this situation. So, to compute $I(\eta_I^A,\eta_I^B)$ from Eq.~\eqref{I integral cosmological expansion pass 1}, we can integrate over $k$, obtaining
    \begin{align}
        I(\eta_I^A,\eta_I^B)=&-\frac{1}{(2\pi)^2a(\eta_I^A)a(\eta_I^B)}\left[\frac{d}{\epsilon}\frac{1}{d^2+\epsilon^2}+\frac{6\xi-1}{2}\right.\nonumber\\&\left.\times\left(\text{atan}(d/\epsilon)\int_{\eta_I^A}^{\eta_I^B}a^2(\eta)H_c^2(\eta)d\eta-\int_{\eta_I^A}^{\eta_I^B}a^2(\eta)H_c^2(\eta)\left(\text{atan}\left(\frac{\eta_I^B-\eta}{\epsilon}\right)-\text{atan}\left(\frac{\eta_I^A-\eta}{\epsilon}\right)\right)d\eta\right)\right]\,.\label{Integral I complete}
    \end{align}
\normalsize

A long distance communication implies $d\gg\epsilon$, so that we can approximate $d^2+\epsilon^2\sim d^2$ and
$\arctan(d/\epsilon)\sim\frac{\pi}{2}$ in the first term and second term of Eq.~\eqref{Integral I complete}, respectively. Analogously, in the third term of Eq.~\eqref{Integral I complete}, the function $\arctan((\eta_I^B-\eta)/\epsilon)-\arctan((\eta_I^A-\eta)/\epsilon)$ can be approximated to $\pi$ whenever $\eta_I^B-\eta\gg\epsilon$ and $\eta-\eta_I^A\gg\epsilon$. By running $\eta$ in the integration range $(\eta_I^A,\eta_I^B)$, the conditions $\eta_I^B-\eta\gg\epsilon$ and $\eta-\eta_I^A\gg\epsilon$ are always satisfied except when $\eta$ is in a neighborhood of width $\sim\epsilon$ of $\eta_I^A$ or of $\eta_I^B$. However, since $d\gg\epsilon$, the contribution of these neighborhoods are negligible in the entire integration range $(\eta_I^A,\eta_I^B)$. Henceforth, since $d\gg\epsilon$, the third term of the integral in equation \eqref{Integral I complete} can be approximated to $\pi\int_{\eta_I^A}^{\eta_I^B}a^2(\eta)H_c^2(\eta)$ and the integral $I(t_I^A,t_I^B)$ in Eq.~\eqref{Integral I complete} can be finally written as
\begin{align}
    I(\eta_A,\eta_B)\sim-\frac{1}{(2\pi)^2a(\eta_I^A)a(\eta_I^B)}\frac{1}{d}\times\left(\frac{1}{\epsilon}-\frac{\pi\left(6\xi-1\right)}{4}\int_{\eta_I^A}^{\eta_I^B}H_c^2(\eta)d\eta\right)\,.
\end{align}\normalsize

For the integrals $J_i(t_I^i)$, by using the decomposition \eqref{normal mode decomposition}, from Eq.~\eqref{J definition}, we have
\begin{equation}\label{J integral first passage}
    J_i(\eta_I^i)=\frac{1}{(2\pi^2)a^2(\eta_I^i)}\int_0^\infty k^2e^{-2k\epsilon}|\chi_k(\eta_I^i)|^2dk\,.
\end{equation}
Using the perturbative solution for $\chi_k$, Eq.~\eqref{chi perturbative}, up to first order, we have \normalsize
\begin{equation}\label{chi modulus square}
    |\chi_k(\eta)|^2\sim\frac{1}{2k}-\frac{\left(6\xi-1\right)}{k^2}\int_{\eta_I^A}^{\eta_I^B}\sin(2k(\eta_I^B-\eta))a^2(\eta)H_c^2(\eta)d\eta\,.
\end{equation}\normalsize
Putting Eq.~\eqref{chi modulus square} into Eq.~\eqref{J integral first passage} we end up with
\begin{align}
    J_i(\eta_I^i)=&\frac{1}{(2\pi)^2a^2(\eta_I^i)}\left(\frac{1}{4\epsilon^2}-\left(6\xi-1\right)   \int_{\eta_I^A}^{\eta_I^i}a^2(\eta)H_c^2(\eta)\frac{\eta_I^i-\eta}{(\eta_I^i-\eta)^2+\epsilon^2}d\eta\right)\,.\label{J integral cosmological expansion}
\end{align}

Interstingly, we can now evaluate the ratio $\frac{I^2(t_I^A,t_I^B)}{J_A(t_I^A)J_B(t_I^B)}$ appearing in Eq. \eqref{maximum transmissivity 2},
\begin{align}\label{ratio CE result}
    \frac{I^2(t_I^A,t_I^B)}{J_A(t_I^A)J_B(t_I^B)}\sim 16\frac{\epsilon^2}{d^2}\left(1+\left(1-6\xi\right)F\right)\,,
\end{align}
where
\begin{align}
    F\coloneqq\int_{\eta_I^A}^{\eta_I^B}\left(\frac{\pi\epsilon}{2}-\frac{4\epsilon^2(\eta_I^B-\eta)}{(\eta_I^B-\eta)^2+\epsilon^2}\right)a^2(\eta)H_c^2(\eta)d\eta\,.\label{factor F}
\end{align}
From Eq.~\eqref{maximum transmissivity 2}, we have
\begin{equation}\label{maximum transmissivity cosmological expansion}
    \tau=\frac{16}{\alpha_A^2}\frac{\epsilon^2}{d^2}(1+(1-6\xi)F)\,.
\end{equation}
To conclude, one can easily see that the relative error \eqref{Perturbation error chi} on $\chi_k(\eta)$ in Eq.~\eqref{chi perturbative} propagates in the transmissivity $\tau$ in Eq.~\eqref{maximum transmissivity cosmological expansion} to become
\begin{equation}
    \mathcal{E}_P\coloneqq 4\int_{\eta_I^A}^\eta\int_{\eta_I^A}^{\eta_1}(\eta-\eta_1)(\eta_1-\eta_2)U(\eta_1)U(\eta_2)d\eta_1d\eta_2\,,\label{Perturbation error}
\end{equation}

\section{Scale factor of the Einstein-de Sitter universe in conformal time}\label{appendix: scale factor}

To solve the Friedmann equation \eqref{Friedmann Eq. perfect fluid} for an EdS universe, it is more convenient to use the cosmic time $t$. In so doing, having $H\coloneqq\frac{\dot{a}}{a}$, Eq.~\eqref{Friedmann Eq. perfect fluid} becomes
\begin{equation}\label{Friedmann Eq. cosmic time}
    \dot{H}+\frac{3(1+w)}{2}H^2=0\,.
\end{equation}
By imposing $a(t_I^A)=1$ one can easily solve Eq.~\eqref{Friedmann Eq. cosmic time} to obtain the scale factor
\begin{equation}\label{scale factor perfect fluid cosmological time}
    a(t)=\left(\frac{t}{t_I^A}\right)^{\frac{2}{3w+3}}\,.
\end{equation}
We now want to use the conformal time $\eta(t)$ satisfying $d\eta=\frac{dt}{a(t)}$. By defining $\eta_I^A\coloneqq \eta(t_I^A)=t_I^A$ we get
\begin{equation}\label{computationStep1}
    \eta-\eta_I^A=\int_{t_I^A}^{t}\left(\frac{t_I^A}{t_1}\right)^{\frac{2}{3(w+1)}}dt_1=(t_I^A)^{\frac{2}{3(w+1)}}\frac{t^{1-\frac{2}{3(w+1)}}-(t_I^A)^{1-\frac{2}{3(w+1)}}}{1-\frac{2}{3(w+1)}}=\frac{3(w+1)}{3w+1}\left((\eta_I^A)^{\frac{2}{3(w+1)}}t^{\frac{3w+1}{3(w+1)}}-\eta_I^A\right)\,.
\end{equation}
From Eq.~\eqref{computationStep1}, we can obtain $t$ in terms of $\eta$ as
\begin{equation}
    t=\left(\frac{3w+1}{3(w+1)}(\eta_I^A)^{-\frac{2}{3(w+1)}}\eta+\frac{2}{3(w+1)}(\eta_I^A)^{\frac{3w+1}{3(w+1)}}\right)^{-\frac{3(w+1)}{3w+1}}\,.
\end{equation}
In particular
\begin{equation}\label{computation step 3}
    \left(\frac{t}{t_I^A}\right)^{\frac{3w+1}{3(w+1)}}=\frac{3w+1}{3(w+1)}\frac{\eta}{\eta_I^A}+\frac{2}{3(w+1)}\,.
\end{equation}
Now, to obtain the scale factor \eqref{scale factor perfect fluid cosmological time}, it is sufficient to raise both sides of Eq.~\eqref{computation step 3} to the power of $\frac{2}{3w+1}$, getting
\begin{equation}\label{scale factor perfect fluid conformal time}
    a(\eta)=\left(\frac{3w+1}{3w+3}\frac{\eta}{\eta_I^A}+\frac{2}{3w+3}\right)^{\frac{2}{3w+1}}\,.
\end{equation}
We can finally compute $aH_c(\eta)=\frac{a'}{a}$. First, we calculate the derivative of $a$ as
\begin{equation}
a'(\eta)=\frac{2}{3(w+1)}\left(\frac{3w+1}{3(w+1)}\frac{\eta}{\eta_I^A}+\frac{2}{3(w+1)}\right)^\frac{1-3w}{1+3w}\,.
\end{equation}
Then
\begin{equation}
    aH_c=\frac{a'}{a}=\frac{2}{3(w+1)\eta_I^A}\left(\frac{3w+1}{3(w+1)}\frac{\eta}{\eta_I^A}+\frac{2}{3(w+1)}\right)^{-1}=\frac{1}{\frac{3w+1}{2}\eta+\eta_I^A}\,,
\end{equation}
obtaining the result reported in Eq.~\eqref{conformal Hubble parameter perfect fluid}.
\end{document}